\documentclass[preprint,12pt]{elsarticle}
\usepackage{amssymb}
\usepackage{amsmath}
\usepackage{latexsym}
\usepackage{lineno}
\usepackage{fancyvrb}
\usepackage{color}
\usepackage[margin={2.5cm,2cm}]{geometry}  
\usepackage[colorlinks=true,linkcolor=blue,urlcolor=blue,citecolor=blue,anchorcolor=blue]{hyperref}

\newcommand{\D}{\partial}

\begin{document}

\begin{frontmatter}

\title{Thermally-driven scintillator flow in the SNO+ neutrino detector}

\bigskip

\author[inst]{J.D. Wilson for the SNO+ Collaboration\corref{cor1}}
\affiliation[inst]{organization={Department of Earth \& Atmospheric Sciences, University of Alberta},
            city={Edmonton},
            state={Alberta},
            country={Canada}}

\cortext[cor1]{Corresponding author: J.D. Wilson (johnd@ualberta.ca, 1 780 909 1969), Dept. Earth \& Atmospheric Sciences, University of Alberta, 1-26 Earth Sciences Building, Edmonton, Alberta, Canada T6G 2E3.}

\pagebreak

\pagenumbering{arabic}
\setlength{\baselineskip}{18pt}

\begin{abstract}

The SNO+ neutrino detector is an acrylic sphere of radius 6 m filled with liquid scintillator, immersed in a water-filled underground cavern, with a thin vertical neck (radius 0.75 m) extending upwards about 7 m from the sphere to a purified nitrogen cover gas. To explain a period of unexpected motion of the scintillator, time-dependent flow simulations have been performed using OpenFoam. It appears that the motion, inferred from subsequent 24 h-averaged patterns of transient radon ($^{222}\mathrm{Rn}$) contamination introduced during earlier recirculation of scintillator, can be explained as owing to heat transfer through the detector wall that induced buoyant flow in a thin wall boundary layer. This mechanism can result in transport of contaminant, should it be introduced, down the neck to the sphere on a time scale of several hours. If the scintillator happens to be thermally stratified, the same forcing produces internal gravity waves in the spherical flow domain, at the Brunt-V\"{a}is\"{a}l\"{a} frequency. Nevertheless, oscillatory motion being by its nature non-diffusive, simulations confirm that imposing strong thermal stratification over the depth of the neck can mitigate mixing due to transient heat fluxes.

\end{abstract}

\begin{highlights}

\item Simulation of thermally-driven scintillator motion in a spherical neutrino detector 
\item Fluid ascends (descends) in laminar boundary layer along relatively warm (cool) wall 
\item Volume conservation demands compensating descent (ascent) of interior fluid
\item This mechanism explains observed motion of transient radon contamination 
\item $\mathcal{O}[0.1 \, \mathrm{W \, m^{-2}}]$ heat flux (or $\mathcal{O}[0.1 \, \mathrm{K}]$ thermal inhomogeneity) $\rightarrow$ speeds $\mathcal{O}[\mathrm{mm \, s^{-1}}]$

\end{highlights}


\end{frontmatter}

\setlength{\baselineskip}{18pt}

\section{Introduction}

This paper will report numerical simulations of laminar convective flow within a spherical container, carried out using OpenFoam to investigate a flow phenomenon observed within the SNO+ liquid scintillator particle detector. It complements several earlier papers \citep{DiMarcello_etal_NIM2020,DiMarcello_etal_JFE2022,ShuZhang_etal2021} addressing a similar subject, and although the novelty of the work lies in its context rather than its methodology, the simplicity of the problem (in terms of domain geometry, fluid homogeneity and thermal forcing) implies that the results we present should have some generality, supplementing what is a surprisingly sparse literature \cite{Pepper_Heinrich1993} on a common and important type of flow \citep{Hutchins_Marschall1989,Arquis_etal1993} occurring in laboratory beakers, storage tanks and so forth.

\begin{figure}[h!]
\begin{center}
\includegraphics[width=10 cm]{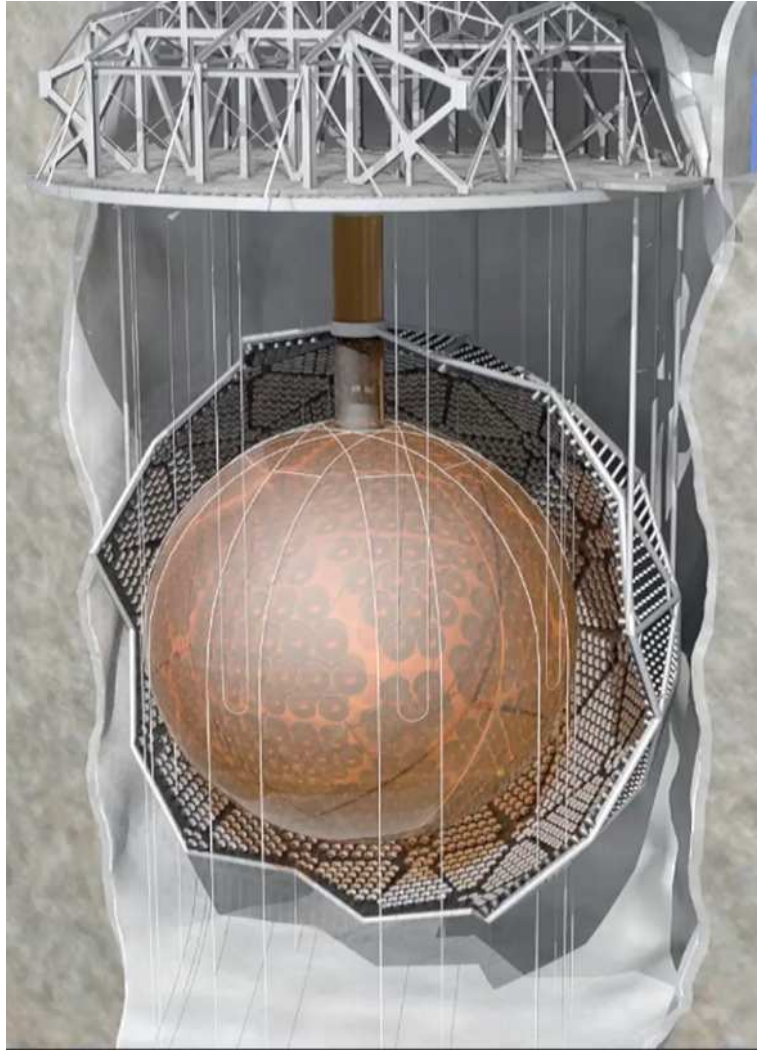}
\end{center}
\caption{Schematic diagram of the SNO+ neutrino detector, the ``flow domain'' of interest \cite{Albanese_etal_SNOplusCollab_JINST2021}. Floating in water in the rock cavity, an acrylic sphere (radius 6 m) is surrounded by the PSUP (PMT support frame) bearing photomultiplier tubes (PMTs). The sphere is connected by the `neck' (radius 0.75 m) to the `deck' some 7 m above the sphere. High in the neck is the liquid/gas interface, where the scintillator meets the purified $\mathrm{N}_2$  `cover gas'.}\label{detector}
\end{figure}

For present purposes there is no need to delve deeply into the SNO+ science goals and methodology (for details see \citep{Albanese_etal_SNOplusCollab_JINST2021,Anderson_SnoplusCollab_2021,Andringa_SnoplusCollabUnnamed_2016}). The SNO+ detector is located in a rock cavity some 2 km underground near Sudbury, Ontario. The scintillator fluid, a linear alkylbenzene (LAB), is contained within an acrylic sphere of 6 m radius (Fig.~\ref{detector}), and rises up a cylindrical `neck' (radius 0.75 m) to the interface with a `cover gas' of ultrapure N$_2$ within the Universal Interface (UI), about 13 m above the centre of the sphere. The detector sphere (hereafter `AV' for `acrylic vessel') and neck `float' within a surrounding mass of purified water that fills the rock cavity to a level about 1 m below the top of the neck, hold-down ropes compensating for the buoyancy of the AV (the scintillator's density is nominally about $860 \, \mathrm{kg \, m^{-3}}$). Within the water, approximately concentric with the AV and at a radius of about 8.5 m from its centre, some 9400 inward-looking photomultipler tubes (PMTs), mounted on a frame referred to as the PSUP (PMT Support), respond to photons emerging from the detector --- and carry the information from which physics data are deduced.

It is characteristic of this type of detector that one seeks to discriminate a relatively small number of target events, e.g. scattering of an electron by a solar neutrino, or a hypothetical neutrinoless double beta decay, from a `background' of vastly more numerous events that stem mostly from natural radioactivity within and near the detector. Accordingly every effort is made to minimize radioactive contamination, by optimally selecting material parts, and by filtration and purification of the scintillator and external water. Nevertheless in practice the scintillator will carry a low level of activity, particularly from short-lived progeny of the $^{238}\mathrm{U}$ and $^{232}\mathrm{Th}$ chains during and after filling and recirculation operations, and this is monitored by quantifying the associated event rates. The contaminant of primary interest in this paper is radon ($^{222}\mathrm{Rn}$, half-life 3.8 days), which is present in the laboratory air at a level of about $10^2 \; \mathrm{Bq \, m^{-3}}$ \cite{Albanese_etal_SNOplusCollab_JINST2021}. Within the AV the concentration of $^{222}\mathrm{Rn}$ can be quantified from the decay rates of its daughter radionuclides, i.e. the transformation of the daughter bismuth ($^{214}\mathrm{Bi}$) by beta decay to polonium ($^{214}\mathrm{Po}$) and the subsequent alpha decay of the $^{214}\mathrm{Po}$, a sequence producing a delayed coincidence signal that is easily identified. It is critical for SNO+ that the radon level in the detector be many orders of magnitude below that in ambient air or water, which is achieved by isolating the scintillator from room air to the maximum possible degree\footnote{At present the ratio of the radon concentration in the cover gas to that in the air is smaller than $2 \times 10^{-4}$ \cite{Albanese_etal_SNOplusCollab_JINST2021}, and further improvement is anticipated. Please note that the level of BiPo-214 activity evident in Figs.~(\ref{blob_Valentina_zrho2_plot},\ref{snoplus31_run30Dec_blob_obs_vs_model_Nr40_Nz36}) below was seen just after the completion of scintillator operations to mix in the fluor (2,5-diphenyhloxazole or `PPO', incorporated at a concentration of 2.2 g/L as a wavelength-shifter), and is much higher than, and {\em not characteristic of}, the level now achieved.}.

Evidently then, in the context of the {\em purity} of liquid scintillator particle detectors --- SNO+ being one of several in operation --- one is going to be concerned to understand and perhaps control scintillator motion, which may transport a contaminant into the ``fiducial volume'', i.e. a subregion of scintillator sufficiently distant from the wall as to exclude background events occurring in or near the AV walls. Absent any pumping or intentional agitation, scintillator motion can be driven only by thermal gradients or temporal trends, and in practice such convective motion fields are very weak. In this regard one may take the example of the Borexino (liquid scintillator) underground neutrino detector, which is similar in size and type to SNO+, and whose scintillator motion patterns have been investigated in a sequence of papers \citep{BravoBerguno_etal2018,DiMarcello_etal_NIM2020,DiMarcello_etal_JFE2022}. Early measurements gave proof of thermally-driven motion within the detector, driven largely by seasonal variation of laboratory air temperature, and modifications were made to reduce the amplitude of that motion (insulation of the outer wall of the detector, and provision of active temperature controls) with the result that lateral inhomogeity of the detector's boundary temperature was reduced to a level of order 0.1 K. Even so, the scientific goals of the Borexino experiment necessitate quantification of the weak convective motion\footnote{These authors note the paucity of existing studies of the ``natural convection problem under consideration, represented by fluid flow inside [a] closed, stratified, near-equilibrium system''.}, as reported by \cite{DiMarcello_etal_JFE2022}.

Returning to the SNO+ experiment, it had been envisaged that other than during intervals of `AV recirculation' (i.e. extraction, purification and reintroduction of LAB), stable thermal stratification of scintillator in the neck would mitigate convective transport from the gas/liquid interface at the top of the neck down into the fiducial volume. In that context, and motivated by observation of an unanticipated pattern of scintillator motion, it is the purpose of this paper to present numerical simulations that illustrate and typify the types of flow that may occur in the detector in response to idealized but plausible thermal inhomogeneities.

The outline of the paper is as follows. Section 2 will detail an actual contamination event, a quiescent period (i.e. no pumping) during which a transient and decaying radon layer was observed to slowly sink within the AV. Section 3 will briefly cover OpenFoam (`OF', the open source software used to simulate flow in the detector) and its method of application. Direct scintillator velocity measurements are unavailable, and so in Section 4 simulations of a closely analogous flow, viz. that within a 3 L spherical flask of water subjected to controlled warming from its boundary, is compared with corresponding measurements \cite{Chow_Akins_1975}. Confidence in the methodology having been established, Section 5 will show that modest wall heat fluxes (of order $0.1 \, \mathrm{W \, m^{-2}}$) induce buoyancy-driven motion of scintillator within the spherical AV that explains (semi-quantitatively) what has been observed, and that if such forcing combines with stable temperature stratification of the scintillator fluid a high frequency wave motion ensues. Finally Section 6 focuses on mixing down the neck, as driven by distinct types of thermal disturbance, and examines the impact of bulk thermal stratification upon that mixing.

\section{Observed motion of a contamination layer}\label{sec:blob}

In early June 2021, just after the LAB fill had been completed, an interesting event was observed in the pattern of BiPo-214 decays within the SNO+ detector. After some 5 h of AV recirculation on 31 May, a radon-contaminated layer had formed\footnote{The $^{222}\mathrm{Rn}$ entered with the AV recirculation operations, as expected and as previously seen by other scintillator experiments. During `normal' recirculation, scintillator is extracted from the bottom of the AV and re-injected at the base of the neck. Experience has shown that the behaviour of the re-introduced scintillator depends on its temperature relative to that within the AV. Newly returned fluid {\em may} promptly undergo mixing, or may form a transient blob or layer at the top of the AV sphere that later sinks.}, rather uniform in its activity (as seen in the pattern formed by daily-summation of BiPo-214 decays) and spanning approximately the layer $2.5 \le z \le 5$ m (the coordinate origin used here is the centre of the AV sphere, and $z$ is the vertical coordinate). During the subsequent two weeks, until the next interval of AV recirculation on 14 June, the AV was subjected to no disturbance other than the nitrogen ($\mathrm{N}_2$) bubblers\footnote{One bubbler emits at the base of the AV, and another at the base of the neck. The bubble stream consists of small -- circa. 1 cm diameter --  single bubbles, with a cumulative volume flow rate (at atmospheric pressure) of about 150 L in two weeks.}, and any unmeasured thermal inhomogeneity or trend.

It was observed that over those two weeks the contaminated layer slowly sank, decaying as expected owing to the short (3.8 day) half-life of $^{222}\mathrm{Rn}$, but more or less retaining its layered form. Fig.~(\ref{blob_Valentina_zrho2_plot}) plots that slow descent, as captured by daily integrations of the Bi-214 and Po-214 decays plotted in $(z,\rho^2)$ coordinates, where $\rho=\sqrt{x^2+y^2}$.

The radon `blob' of 31 May - 14 June 2021 was not the first such to be observed, but it stands out because of its occurrence during a long interval without external disturbance (the normal situation was for recirculation operations to be paused only on weekends). The need to understand the origin of these events and their evolution prompted the present work, a numerical study of thermally-driven circulation in the AV. What is sought is an initial thermal state, and a choice of thermal forcing on the AV wall, that results in the (assumed) static initial state developing a convective circulation that achieves the observed `translation' of the blob.

\subsection{Further features of `the blob'}

The volume of scintillator recirculated on 31 May was $\sim 17 \mathrm{m^3}$, more than the  volume of the entire neck, which is $V_{\mathrm{neck}} \sim \pi (0.75)^2 \times 6.8 \sim 12 \; \mathrm{m^3}$. However the initial volume of the contaminated layer, estimated as
\begin{equation}
V_{\mathrm{blob}} \approx  \int\limits_{z=2.5}^5 \; \pi \rho^2(z) \; dz \; ,
\end{equation}
is $\sim 170 \; \mathrm{m^3}$. This implies that the `blob' of 31 May was not an undiluted mass of recently recirculated scintillator --- mixing had already occurred.

\begin{figure}[h!]
\begin{center}
\includegraphics[width=17 cm]{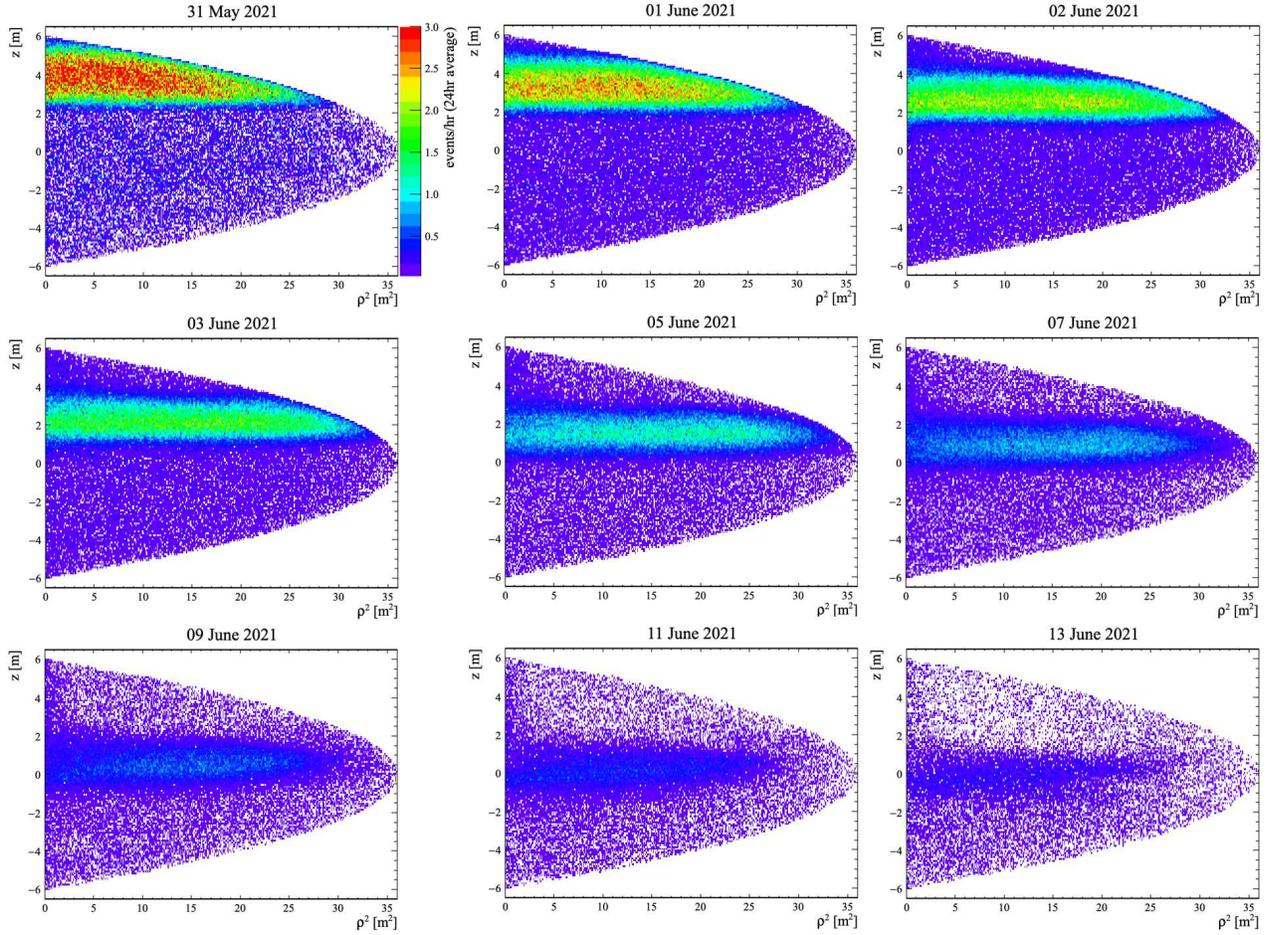}
\end{center}
\caption{The sinking $^{222}\mathrm{Rn}$ layer of 31 May - 14 June, 2021, as seen in daily-summed event distributions (sum of $^{214}\mathrm{Bi}$ and $^{214}\mathrm{Po}$ decays) plotted in $(z, \rho^2)$ coordinates ($\rho=\sqrt{x^2+y^2}$). The colour code gives the hourly event count per $z-\rho^2$ bin, averaged over 24 hr (bin widths $\Delta z, \Delta \rho^2$ respectively $0.1 \, \mathrm{m}$ and $0.1 \, \mathrm{m^2}$). Note: the high level of contamination indicated here is not typical of the SNO+ detector, and represents an anomaly.}\label{blob_Valentina_zrho2_plot}
\end{figure}

By inspection of Fig.~(\ref{blob_Valentina_zrho2_plot}) a descent rate of the order of $1 \, \mathrm{cm \, hr^{-1}}$ (or $1/4 \; \mathrm{m \, day^{-1}}$) can be inferred for the contaminated layer, decreasing as time increases. If Fig.~(\ref{blob_Valentina_zrho2_plot}) is interpreted as showing a layer that simply sinks, undistorted, in the AV, there must be a mechanism to move uncontaminated fluid from beneath to above the blob. Letting $z_m$ be the mean height of the blob and supposing that the volumetric flow rate of uncontaminated fluid across that plane is $Q(z_m) \; \mathrm{m^3 \, s^{-1}}$, then in the simplest, purely geometrical picture the descent rate is
\begin{equation}
\frac{d z_m}{dt} \approx \frac{Q(z_m)}{\pi} \; \frac{1}{z_m^2-R^2} \, . 
\end{equation}
For example if $z_m=4$ m and $dz_m/dt=-0.25/(24 \times 3600)$, i.e. 0.25 m sink over one day, then $Q \sim \tfrac{1}{2} \, \mathrm{m^3 \, hr^{-1}}$.

Unfortunately the thermal stratification of the scintillator, relevant to any understanding of its motion, is not measured. A vertical profile of temperature is however measured in the cavity water,  some metres outside the PSUP. As of 3 June 2021 that profile had been rather steady over the preceding weeks, and was height-invariant to within about $\pm 0.5$ K from the base of the rock cavity to a height just a few metres below the base of the neck. If one assumes the temperature profile within the AV was qualitatively similar, then the descent of the blob within the AV occurred in a thermal environment that did not mitigate firmly against vertical motion, except near the top of the neck, a region that should be irrelevant to motion within the AV sphere.

\section{OpenFoam}

Calculations were performed using {\em OpenFoam} (`OF') v2006\footnote{From OpenCFD (\href{https://www.openfoam.com/}{www.openfoam.com}), an affiliate of Engineering Systems International (ESI).}, running under Linux (OpenSuSe Leap 15.2 in single processor mode, or Ubuntu 20.04 in multi-processor mode). The surface shell for each chosen flow domain was generated using {\em Salome}\footnote{From Open Cascade, \href{https://www.salome-platform.org/}{www.salome-platform.org}} to produce `stl' (surface triangle language) files, from which OF's blockMesh and snappyHexMesh utilities generated the three-dimensional computational mesh\footnote{The mesh cells are polyhedra, the majority (more specifically) hexahedra.}. Far from domain boundaries the cell size  was uniform, and controlled by a setting in OF's `blockMeshDict' file. Near the walls, layers of cells were added (following a prescription set in `snappyHexMeshDict') to enhance resolution.

The simulations described here used OF's solver for time-dependent compressible flows, `buoyantPimpleFoam', with an added field variable $C$ representing $^{222}\mathrm{Rn}$ and obeying the conservation equation
\begin{equation}\label{eq:massconsv}
\frac{\D \rho C}{\D t} = \, - \, \nabla \cdot \left(\rho \mathbf{U} C \, - \, \rho \mathcal{D}_c \, \nabla C \right) \, - \, \rho \, \frac{C}{\tau}
\end{equation}
where $\mathbf{U}$ is the velocity vector; $\rho$ is the fluid density, modelled\footnote{In this approximation density is independent of pressure.  Some simulations, not reported here, used the `incompressible' solver `buoyantBoussinesqPimpleFoam' (of OF v2006), albeit with an advection term added to the energy equation to parameterize the influence of an implicit uniform and unvarying `background' temperature stratification.} as
\begin{equation}\label{eq:density}
\rho = \rho_0 \, \left[ 1 \, + \, \beta_T \, \left(T-T_0 \right) \right]
\end{equation}
where $\beta_T$ is the thermal expansion coefficient; $\mathcal{D}_c$ is the molecular diffusivity of $C$ (prescribed as $\mathcal{D}_c=\nu/\mathrm{Sc}$, Sc being the Schmidt number); and $\tau$ is the decay lifetime, in the case of $^{222}\mathrm{Rn}$ equal to $4.76 \times 10^5$ s. Tested in an unforced simulation whose strongly stable stratification mitigated against evolution away from a static initial state, the Rn layer, without moving, simply decayed with the imposed half-life. 

Apart from the added field variable $C$, the only other modification made to the solver was to incorporate the Coriolis effect. All simulations assumed laminar flow, which amounts to an assumption that the flows were adequately resolved, both spatially (on the given mesh) and temporally (with the time step applied), obviating any necessity to account for subgrid motion. Further details will be provided below, where they are specific to a given simulation and/or not necessarily obvious even to readers who may be familiar with OpenFoam.

\section{Verification against a measured `analog' flow}\label{sec:ChowAkins}

Chow and Akins \cite{Chow_Akins_1975} reported an experiment in which water, contained in a 3 L spherical flask\footnote{The neck of the flask was inclined at 45$^{\circ}$ to the vertical. For the calculations here, the neck was neglected and the radius $R$ of the sphere computed (from the 3 L volume) to be $R=8.95$ cm.}, was subjected to controlled heating. The flask was immersed within a continuously-stirred water bath, the entire system initially being isothermal at $5^{\circ}$ C. The temperature of the water bath was then progressively increased in such a manner as to sustain a time-invariant temperature difference $T_{\mathrm{b}} - T_{\mathrm{c}}$ between the water bath and the centre of the flask. Suspended in the flask water were glass spheres, whose displacements were measured by a photographic system arranged to provide information on a vertical plane through the centre of the sphere. The authors reported that this fluid system evolves into a pseudo-steady state (i.e. the motion is steady, though the temperature field steadily warms), and presented a single transect of the vertical velocity of the water along an equatorial radius corresponding to the condition that $\Delta T = \overline{T}_{\mathrm{w}} - T_{\mathrm{c}} = 2.5$ K, where $\overline{T}_{\mathrm{w}}$ denotes the {\em mean} temperature of the inner wall, computed by the authors from measurements and known physical properties\footnote{It is {\em possible} that the $2.5^{\circ}\mathrm{C}$ temperature difference had been intended by the authors to refer to the control variable $T_{\mathrm{b}} - T_{\mathrm{c}}$, however certainly they (Chow \& Akins) define $\Delta T$ as ``fluid temperature difference between inside wall and center of sphere'', and they state in the caption for their Fig.~8 that $\Delta T=2.5^{\circ}\mathrm{C}$.}. The value of the Rayleigh number
\begin{equation}
\mathrm{Ra}= \frac{g \, \beta_T \, R^3 \, \Delta T} {\nu \, D_T}
\end{equation}
corresponding to the measured velocity transect was $\mathrm{Ra}=6.5 \times 10^6$ ($D_T$ being the thermal diffusivity; other variables as defined above), and from variations of their experimental configuration Chow and Akins reported that the flow in the flask was laminar provided $\mathrm{Ra} < 10^7$.

Chow \& Akins interpreted (and perhaps designed) their experiment in the context of a prior numerical study \cite{Whitley_Vachon1972} of free convection in a sphere, and their measurement provides here the opportunity to gain confidence in the application of OF, in the context of a flow similar in its geometry and forcing to the case of the SNO+ detector: simulations of the laboratory beaker  experiment and of the SNO+ neutrino detector differ only as regards the {\em length scale} of the mesh, and the physical properties of the fluid. Regarding the latter, Table ~(\ref{tab:thermoType_in_thermophysicalProperties}) specifies the OF thermophysical model used in {\em all} the simulations of this paper. Temperature-dependent properties (density, specific heat, viscosity, Prandtl no.) are represented by polynomials, and for the laboratory beaker simulations, the medium being water, polynomials provided by \cite{HolzmannCFD} (\href{https://holzmann-cfd.com/community/blog-and-tools/cae-blog/thermophysical-properties-water}{HolzmannCFD}) were used.

The domain boundary for simulations of the Chow-Akins flow consisted of a single spherical `patch' (arbitrarily named `shell'), and the desired initial and boundary conditions for temperature were imposed in the OF file `caseFolder/0/T' per the specification of Table~(\ref{tab:init_and_bconds_ChowAkins}). By way of explanation, folder `0' at top level within the `case folder' contains a file specific to each dependent variable, and that file prescribes the initial and boundary conditions in standard OF parlance. It is worth noting that, because the simulation holds the entire surface of the domain boundary at $T_{\mathrm{c}}+2.5^{\circ}\mathrm{C}$, it cannot exactly parallel the experiment, for in the latter the flow itself (or rather, the experimental setup in its entirety) determined the inner wall temperature, which could not have been {\em exactly} isothermal owing to the anisotropy imposed by gravity.

\begin{figure}[h!]
\begin{center}
\includegraphics[width=12 cm]{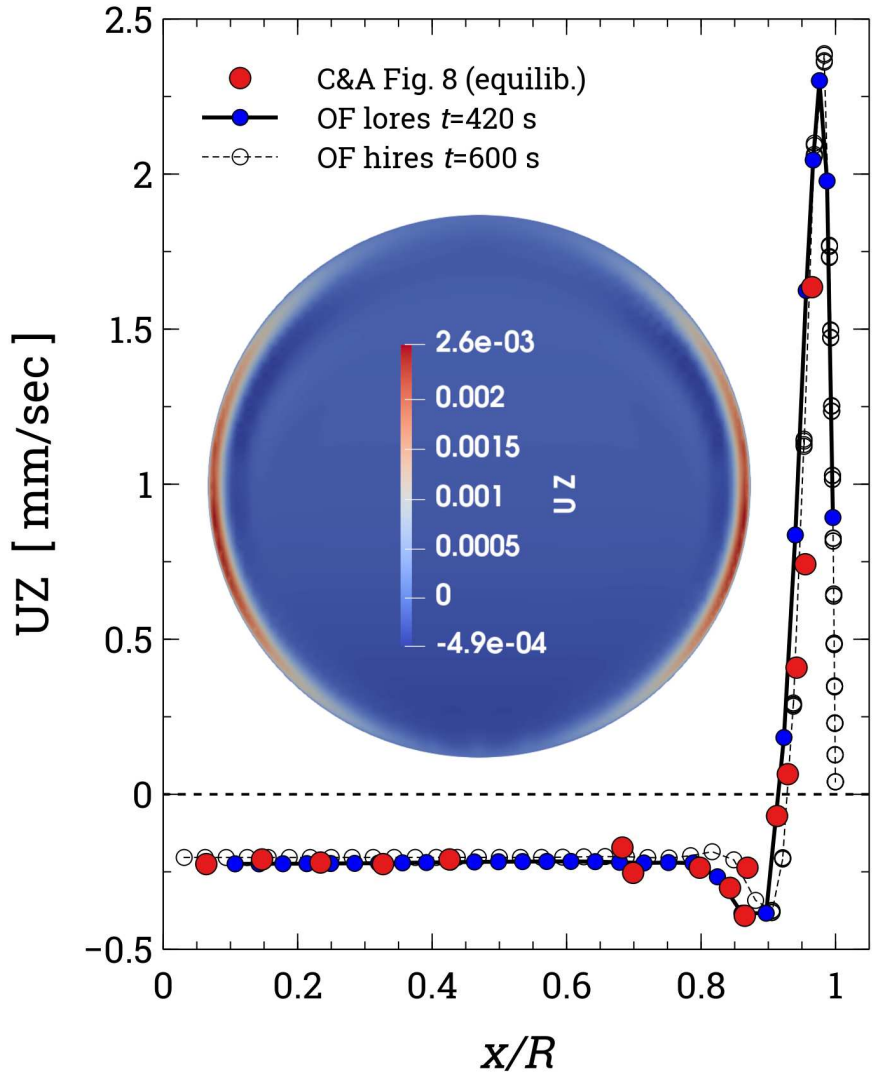}
\end{center}
\caption{Pseudo-steady convective vertical motion of water undergoing heating in a spherical container (radius $R=0.08947$ m). Lineplot: vertical velocity [$\mathrm{mm \, s^{-1}}$] along a radius at the equator, comparing the measurements by Chow \& Akins \cite{Chow_Akins_1975} (Fig. 8) against lower- and higher-resolution OpenFoam simulations. The multiplicity (in the case of the higher resolution simulation) results from the availability of multiple mesh cells in close proximity to the chosen radius ($y=0$, $0 \le x \le R$). Inset: slice $y=0$ of the higher-resolution OF solution at $t=600$ s, with auto-ranged colour scale quantified in $\mathrm{m \, s^{-1}}$. }\label{chow_akins}
\end{figure}

\begin{figure}[h!]
\begin{center}
\includegraphics[width=8 cm]{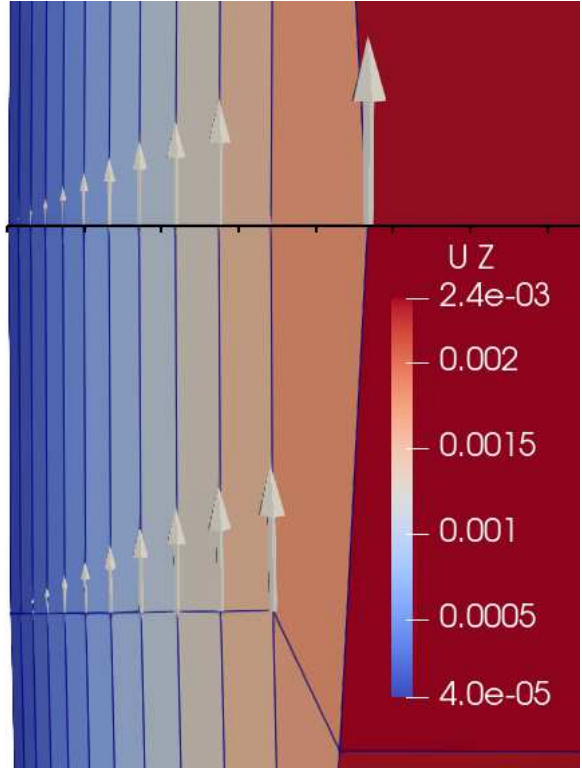}
\end{center}
\caption{A clip of the UZ (vertical velocity) field on $y=0$ at $t=600$ s, from the higher-resolution OF simulation of the Chow \& Akins experiment. The ruler (black) sits along $y=z=0$ with its (left) end at the wall ($x=-0.0895$ m), its gradation interval being 0.1 mm. The ten `added layers' of the mesh are visible, the depth of the outermost layer being about 0.01 mm ($\sim 10^{-4} \times R$). The colour scale range is autoscaled to the visible clip, and gives UZ in $\mathrm{m \, s^{-1}}$.}\label{chow_akins_wall}
\end{figure}

\begin{figure}[h!]
\begin{center}
\includegraphics[width=16 cm]{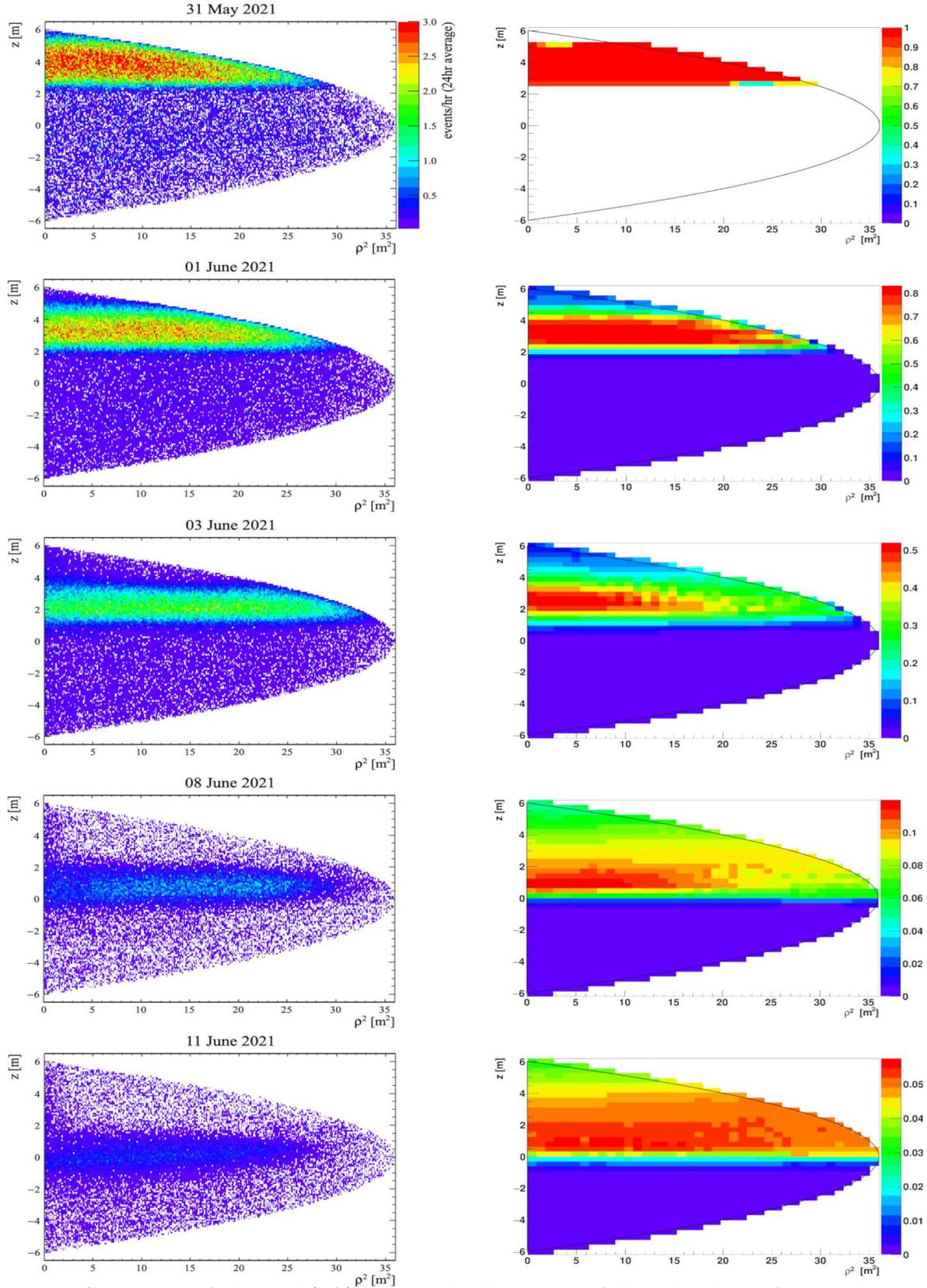}
\vspace{-0.5 cm}\caption{Comparison of observed (left) and simulated positions of the radon layer of 31 May -- 14 June 2021. Axis ranges are $-6 \le z \le 6$ m (vertical) and $0 \le x^2+y^2 \le 36$ m,  (horizontal). Initial state isothermal and static, with concentration $C=1$ over $2.5 \le z \le 5$ m and $C=0$ elsewhere. Evolution forced by heat flux density $q=+0.1 \; \mathrm{W \, m^{-2}}$ on the upper hemisphere. If radon evolution were due solely to radioactive decay, for the times shown (at right) the normalised concentration time sequence would be: $C(t)/C(0)=(1, 0.83, 0.58, 0.23, 0.14)$.}\label{snoplus31_run30Dec_blob_obs_vs_model_Nr40_Nz36}
\end{center}
\end{figure}

Fig.~(\ref{chow_akins}) compares the measured Chow-Akins velocity transect with two simulations:
\begin{itemize}
 
 \item Simulation `lores' was performed on a mesh having a total of 309,290 cells, including those within three layers of cells impressed (by the OF utility snappyHexMesh) to represent the near wall region. A 1st-order accurate discretization (`Gauss upwind') was used for convective terms, i.e. terms of form $\nabla \cdot (\mathbf{U} \phi)$ where $\mathbf{U}$ is the velocity, with components (UX,UY,UZ) along axes (X,Y,Z) in OF notation, and $\phi$ stands for the convected property. 
 
 \item Simulation `hires' used a mesh with 779,506 cells, there being 10 cell layers impressed to cover the wall region, the outermost of which had a depth about 0.01 mm (i.e. $\sim R/10^4$). For this simulation a 2nd-order accurate discretization was used for convective terms.
\end{itemize}
Overall Fig.~(\ref{chow_akins}) shows that the agreement of the two solutions with each other and with the measurements is very good. On their graph, Chow \& Akins extrapolated from their outermost (largest $R$) measurement back towards zero, implying they had not measured velocities closer to the wall than that outermost datum. Thus it need not be thought that there is a discrepancy between the simulations and the data between the outermost measured velocity and the wall. As is evident from Fig.~(\ref{chow_akins_wall}), showing the simulated velocity profile near the domain wall, certainly the higher-resolution OF run provided good resolution of the near wall flow, and it compares nicely with a solution (for this same problem) given by \cite{Hutchins_Marschall1989}, who numerically integrated the non-dimensional vorticity and temperature equations in polar coordinates, pre-supposing azimuthal symmetry. From Fig.~(\ref{chow_akins_wall}) one sees that the boundary-layer Reynolds number $\mathrm{Re}=U d/\nu$ (where $d\sim 0.5$ mm is the wall boundary layer depth, $U \sim 2 \; \mathrm{mm \, s^{-1}}$ the `free stream' velocity and $\nu \sim 10^{-6} \; \mathrm{m^2 \, s^{-1}}$ the kinematic viscosity) is $\mathrm{O}[1]$, compatible with laminar boundary-layer motion.

Given the near equivalence of the OF model and mesh configurations needed to simulate this small scale experiment and the SNO+ detector, the results of this section provide reassurance that OF should be a satisfactory tool to examine thermally-driven SNO+ flows\footnote{Two provisos must be listed here: (i) that an adequate computational mesh be provided, and (ii) that the distinction in length scale (between SNO+ and the Chow-Akins experiment), viz. a factor of $6/0.09=67$, is not so large as to imply that the regimes of flow in the two cases (under the given type of forcing) are qualitatively dissimilar. According to the simulations, boundary layer Reynolds numbers are of the order of unity for both systems.}. In that regard, and anticipating later results, it is interesting to note that (according to the simulation of the Chow-Akins experiment) the warm wall boundary layer does not `detach' from the wall over the lower hemisphere, and similarly in computations of a `reverse Chow-Akins flow' (i.e. with the wall being held colder rather than warmer than the centre of the flask) the cool wall boundary layer was found to remain attached to the wall over the upper hemisphere.

\section{Simulation of a sinking radon layer}\label{sec:simulatesinkingradonlayer}

Simulations of this section relate to the sinking radon layer of Sec.~(\ref{sec:blob}), with the domain of the calculations encompassing the entire AV, including the neck. The rate of sink (of the 24-h averaged pattern) was of the order of 0.25 m/day, and the mechanism seems likely to have been the transfer of fluid volume within the wall boundary layer from beneath to above the contamination layer. It is of interest, then, to determine an initial state and forcing that will replicate what had been observed. As no direct measurements of motion in the SNO+ AV exist, and the true initial state and thermal boundary conditions are inaccessible, the purpose of the following simulations is not to gain a highly accurate solution for a specific (but unmeasured) case, but rather, to outline what {\em qualitative} patterns of motion exist -- and their qualitative effect as demonstrated by the displacement and mixing of a contamination layer from its initial position.

The puzzle presented by the sinking radon layer (Fig.~\ref{blob_Valentina_zrho2_plot}) was the absence of any apparent transport mechanism, i.e. in particular the level base of the sinking layer seemed to suggest either that the 24h-averaged pattern was hiding an eddy motion that was accomplishing the transport, or, that fluid volume was being invisibly transferred across/through the contaminant layer along the AV axis or walls. A clue as to the nature of the flow forcing was provided by the observation that in early June 2021, i.e. during the phenomenon, the LAB/cover-gas interface was rising. This could be explained only by thermal expansion of the LAB and demanded an effective heat addition rate of about 300 W, a figure that implies a mean heat flux density over the AV wall of the order of $1 \; \mathrm{W \, m^{-2}}$.

Table (\ref{tab:properties}) lists the values adopted for the material properties of LAB. Assuming the motion patterns of interest will primarily be thermally-driven, the property that stands out as being of most importance is the thermal expansion coefficient ($\beta_T$). The most uncertain of the properties listed is the Schmidt number (ratio of kinematic viscosity to mass diffusivity), used to evaluate the mass diffusivity $\mathcal{D}_{\mathrm{c}}$ of a species $C$ whose transport equation (Eq.~\ref{eq:massconsv}) was added to the solver to represent movement and decay of $^{222}\mathrm{Rn}$. This however has no bearing on the computed pattern of {\em flow}, and unless the Peclet number $U \ell/\mathcal{D}_{\mathrm{c}}$ were spectacularly small,  convective transport by the bulk velocity field will be overwhelmingly more imortant than diffusion. The value adopted for the kinematic viscosity $\nu$ is nominal (note: $U$ and $\ell$ are characteristic velocity and length scales).

Within the folder `casefolder/constant' a file (typically named `thermophysicalProperties', but for some solvers named `transportProperties') provides the numerical values of these material properties, along with the user's choices for the equation of state and the thermodynamic energy variable to be adopted (enthalpy or sensible energy). Table~(\ref{tab:mixture_in_thermophysicalProperties_SNO+}) gives the `mixture' block of the OF `thermophysicalProperties' dictionary, for the SNO+ simulations to be shown.

Unless stated otherwise the simulations to follow were made on what will be termed the `medium mesh,' comprising a total of 661,536 cells, and including three layers of cells adjacent to the walls. The depth of the outermost layer of cells was 2.7 cm, and gridlengths based on minimum, mean and maximum cell volume ($V$) were $h=V^{1/3} = (0.009,0.11,0.28) \; \mathrm{m}$. The radon contamination layer initially spanned $2.5 \le z \le 5$ m, with $C=1$ within the layer and $C=0$ elsewhere. The wall boundary condition for $C$, `zeroGradient', assured there would be no flux to or from walls. (Note: $C$ is named `Conc' on some of the figures to follow.) 

\subsection{Initial state isothermal}

It was quickly found that if the wall heat flux were specified to be as large as $q=1 \; \mathrm{W \, m^{-2}}$ the rate of sink of the radon layer was excessively large. However when the forcing heat flux was reduced to $q=+0.1 \; \mathrm{W \, m^{-2}}$ and applied over the upper hemisphere of the AV (with $q=0$ elsewhere), qualitative agreement with the observed behaviour of the radon layer was obtained. This can be seen on Fig.~(\ref{snoplus31_run30Dec_blob_obs_vs_model_Nr40_Nz36}), where the observed (daily-average) contaminant field on several days is compared with the instantaneous contaminant field from the simulation\footnote{For comparison with the detector data the simulated concentration fields have been numerically integrated in azimuth, with resolution $dz \times d(x^2+y^2) = 0.1 \, \mathrm{m} \times 0.1  \, \mathrm{m^2}$. Resolution of the simulated field is limited by the OpenFoam mesh length, which away from the AV wall is approximately $0.25$ m.} at a comparable time since cessation of recirculation. Although the correlation of the respective times is inexact, the strong similarity of observation and simulation suggests the latter has captured the essence of the transport mechanism.

\begin{figure}[h!]
\begin{center}
\includegraphics[width=16 cm]{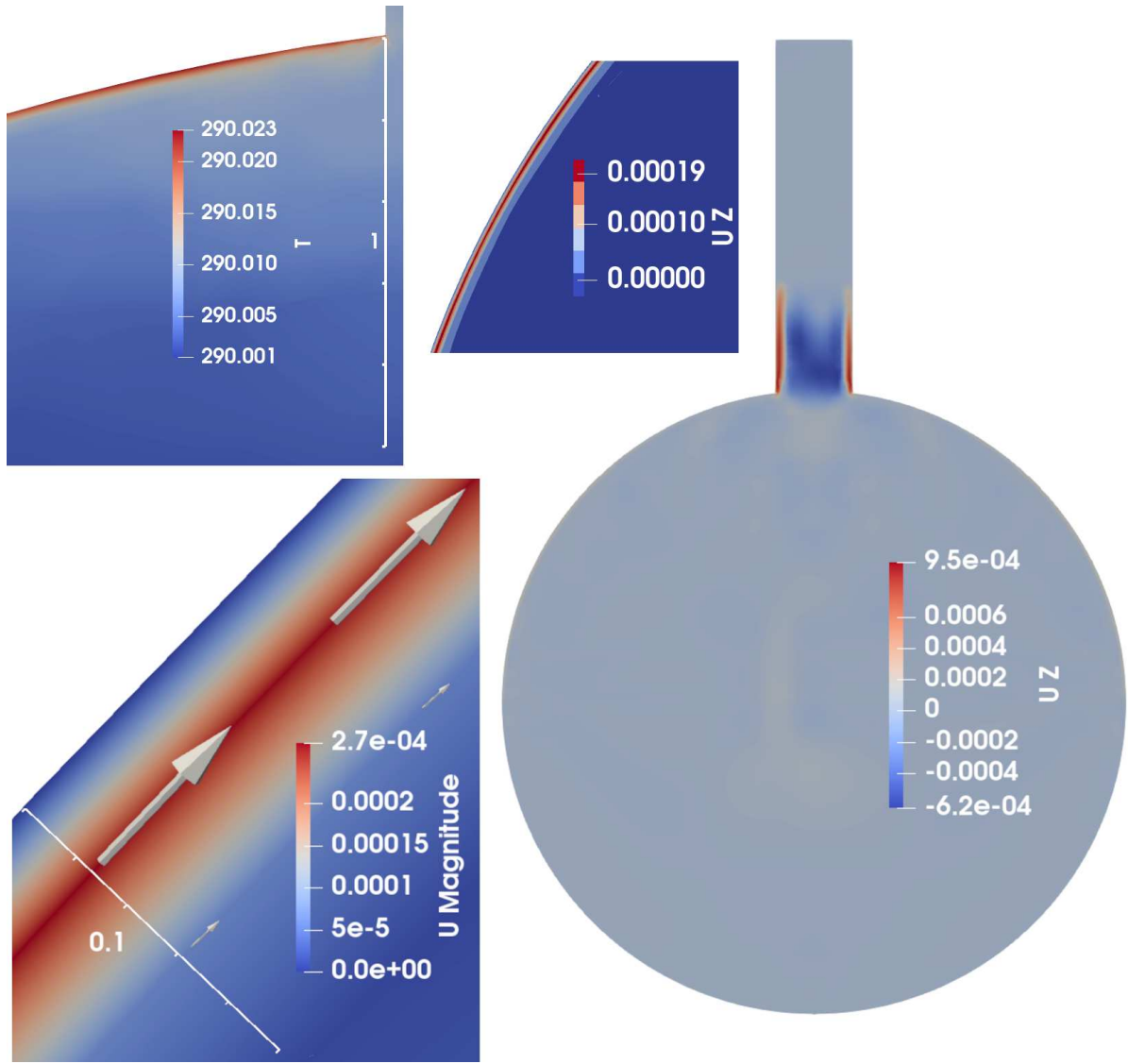}
\end{center}
\caption{Views of the motion in the AV at time $t=12$ h, an isothermal initial state being disturbed by heating ($q=0.1 \; \mathrm{W \, m^{-2}}$) over the the upper hemisphere (excluding the neck). All images are on the $y=0$ slice. Upper-left panel ($T$) shows the warm wall boundary layer (vertical ruler of length 1 m with 0.2 m increments). Upper-middle panel shows vertical velocity (UZ) near the AV wall at $z=3$ m, with (only) 6 colour-values permitted and the scale chosen to distinguish positive from negative UZ (in reality, UZ is {\em not uniform} outside the wall layer). Lower-left panel (velocity magnitude $\mathrm{U}$, $\mathrm{m \, s^{-1}}$) is a blow-up of the wall layer near $z=4$ m (ruler is 0.1 m long). The panel on the right (vertical velocity) shows that the warm wall layer penetrates up along the wall of the neck, inducing sink in the interior of the neck.}\label{snoplus31_run30Dec_combo}
\end{figure}

\begin{figure}[h!]
\begin{center}
\includegraphics[width=14 cm]{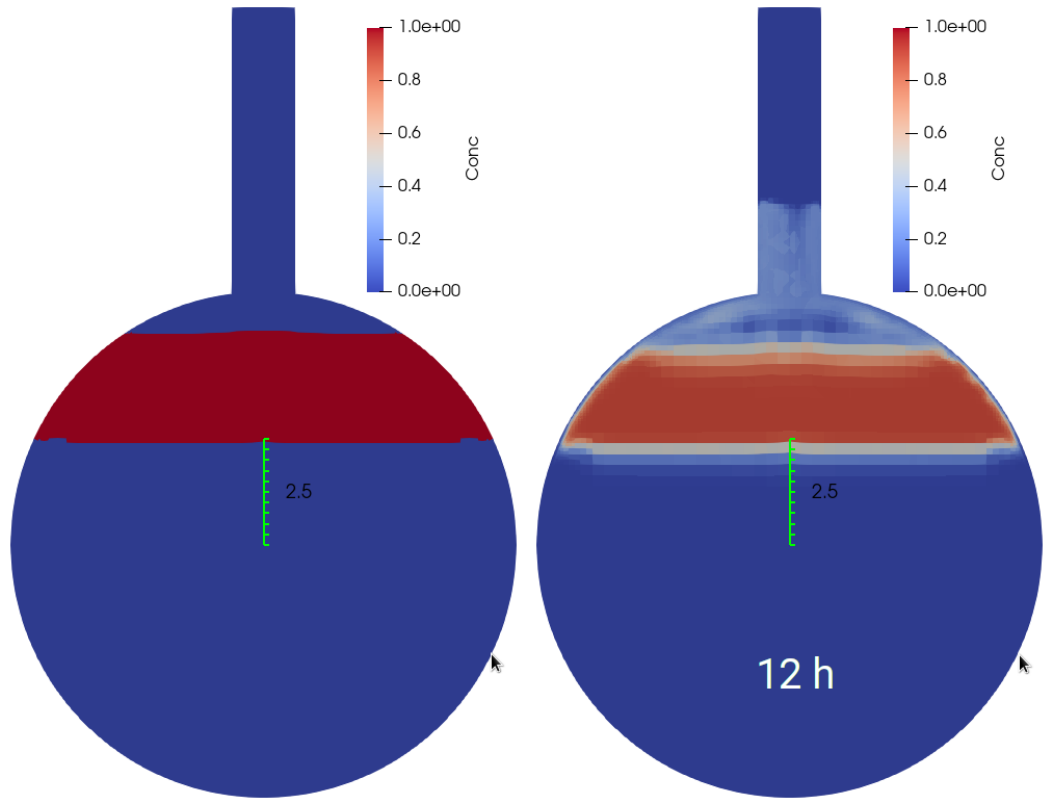}
\end{center}
\caption{The contaminant distribution on $y=0$ at $t=(0,12)$ h. Initial state isothermal, driven by $q=0.1 \; \mathrm{W \, m^{-2}}$ on the upper hemisphere. Ruler is 2.5 m long, with its base at the origin and with 0.25 m increments. Cell sidelength in the interior is about 0.25 m. The ascending warm wall boundary layer is transporting {\em uncontaminated fluid} upwards from beneath the contamination layer, and the accumulation of that fluid volume {\em above} the original contaminant layer forces the latter downward.}\label{snoplus31_run30Dec_conc_0_12h}
\end{figure}

Fig.~(\ref{snoplus31_run30Dec_combo}) summarizes the main qualitative elements of the motion at $t=12$ h after initialization. Buoyancy of the wall boundary layer drives the motion, and away from the wall a compensating counter-current is set up\footnote{This counter-current is obvious to see in the lower neck, on Fig.~(\ref{snoplus31_run30Dec_combo}), but difficult to distinguish in a view of the bulk of the spherical AV, where even in the upper hemisphere it is weak and non-uniform.}, with the result of displacing the contamination layer downward (Fig.~\ref{snoplus31_run30Dec_conc_0_12h}). Adopting from Fig.~(\ref{snoplus31_run30Dec_combo}) approximate values for maximum speed and boundary-layer depth, the boundary-layer Reynolds number $\mathrm{Re}=U d/\nu$ evaluates to $\mathrm{Re} \sim 1$ or smaller -- implying laminar motion.

\begin{figure}[h!]
\begin{center}
\includegraphics[width=8 cm]{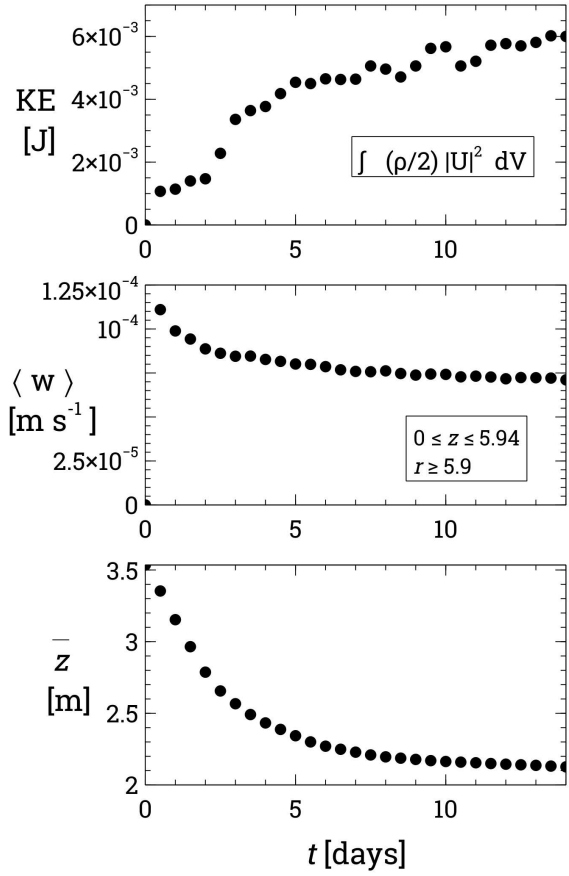}
\end{center}
\caption{Time evolution of total kinetic energy (KE), of the mean vertical velocity ($\langle w \rangle$) computed over a thin outer layer of the upper hemisphere ($0 \le z \le 5.94$ m, $\sqrt{x^2+y^2+z^2}\ge 5.9$ m), and of the contaminant mass weighted mean height of the radon contamination layer (computed over all cells). Initial state isothermal, driven by $q=0.1 \; \mathrm{W \, m^{-2}}$ on the upper hemisphere.}\label{snoplus31_run30Dec_KE_wbar_zbar}
\end{figure}

Fig.~(\ref{snoplus31_run30Dec_KE_wbar_zbar}) plots the time evolution to $t=14$ days of three diagnostics of fluid and contaminant motion. Total kinetic energy is still increasing as of 14 days, albeit at a falling rate, while a measure $\langle w \rangle$ of mean vertical velocity near the upper hemisphere wall (proportional to the ascending volume flux in the wall boundary layer) approaches a steady value. The rate of change of the contaminant-mass weighted mean height
\begin{equation}\label{eq:zbar}
\overline{z} =  \frac{\sum_i \, z_i \, C_i \, dV_i}{\sum_i \, C_i \, dV_i}    
\end{equation}
(where $i$ indexes the cell height $z_i$, concentration $C_i$ and volume $dV_i$) also relaxes with increasing time, as expected. It is worth noting here that due to the specified thermal boundary condition, the simulated flow cannot attain a true steady state: mean temperature must continue increasing, and this impacts the density and consequently (though perhaps to a minor extent) the motion field.

\begin{figure}[h!]
\begin{center}
\includegraphics[width=10 cm]{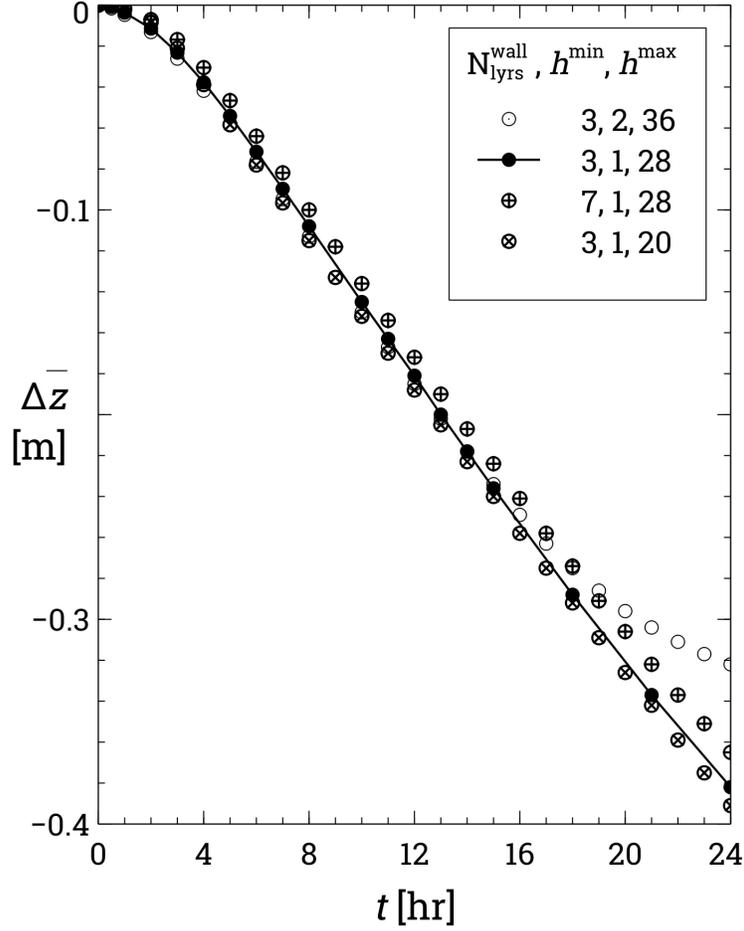}
\end{center}
\caption{Impact of the computational mesh on the rate of descent of the radon contamination layer, for four different choices of mesh (cell counts respectively: $\circ \; 379738$, $\bullet \; 661536$, $\oplus \; 1003934$, $\otimes \; 1088629$). The minimum and maximum cell sizes ($h^{\mathrm{min}}, h^{\mathrm{max}}$) are computed from cell volumes as $V^{1/3}$, and cited in centimeters (rounded). The property plotted is the {\em deviation} ($\Delta \overline{z}$) of the contaminant-mass-weighted mean height of the radon layer from its initial value (the deviation is plotted because the initial value $\overline{z}(0)$ differs slightly from mesh to mesh, owing to the spatial irregularity of the mesh cells produced by snappyHexMesh). In every case an isothermal initial state is disturbed by a heat flux $q=+0.1 \; \mathrm{W \, m^{-2}}$ over the upper hemisphere of the AV, and the OF solver is the same for all cases (i.e. a slight modification of `buoyantPimpleFoam'). The solid circles (and line) correspond to the `standard' mesh used for most `full detector' calculations.}\label{gridindependence}
\end{figure}

Finally, the adequacy of the mesh resolution needs to be addressed. Fig.~(\ref{gridindependence}) compares the rate of descent $d\overline{z}/dt$ of the contamination layer as computed on four different meshes, and suggests that the medium mesh chosen for the simulations of this section is adequate, given the exploratory nature of the investigation.

\subsection{Influence of stratification}

The previous section established a plausible mechanism for the phenomenon that prompted the study, i.e. the observed slow sink (without apparent mixing) of a contaminant layer within the SNO+ neutrino detector. Given the impossibility of knowing the {\em true} initial state and forcing for the June 2021 contamination layer, there is no logical basis for demanding a better accord with the data than has been shown. It remains interesting, however, to determine (again, qualitatively) what effect thermal stratification of the detector might have.

\begin{figure}[h!]
\begin{center}
\includegraphics[width=10 cm]{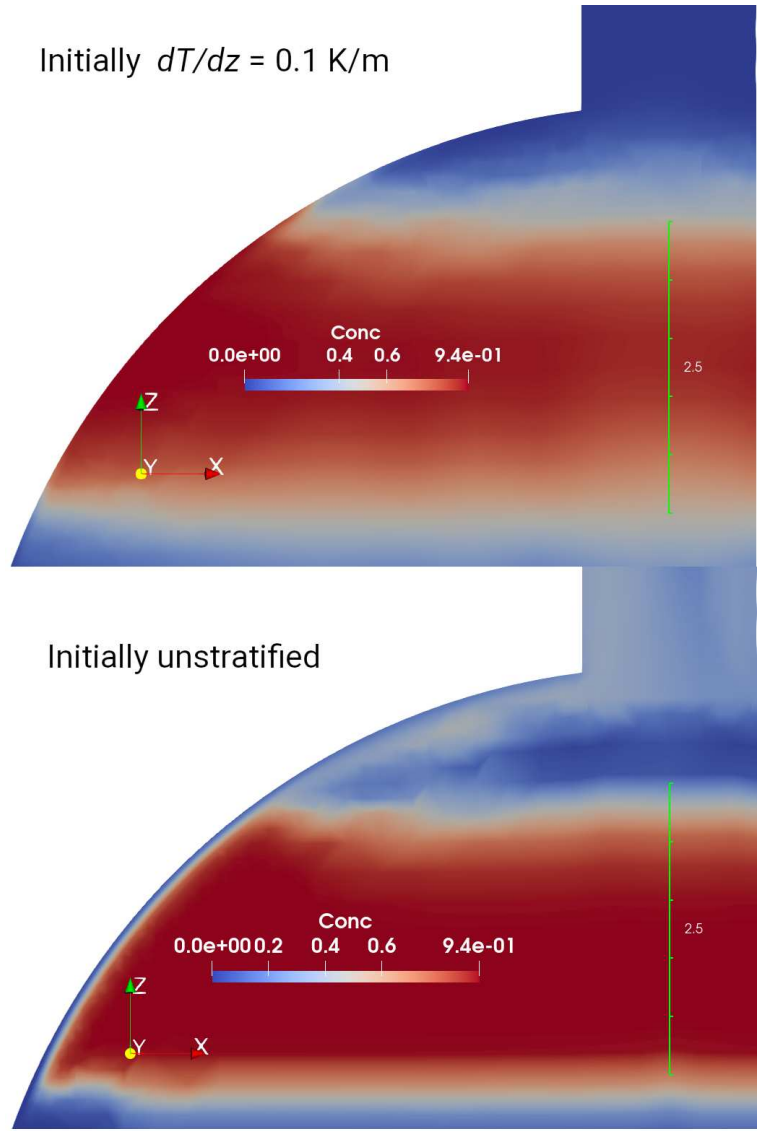}
\end{center}
\caption{Comparison of contaminant concentration fields on $y=0$ at $t=8$ hr for two runs, both forced by heat flux $q=+0.1 \; \mathrm{W \, m^{-2}}$ on the upper hemisphere of the AV. Upper: initial state uniformly stratified, $T=290 + 0.1 (z-5)$. Lower: initial state isothermal. In the isothermal case the wall boundary layer features {\em uncontaminated} fluid, brought up from beneath the radon layer by the buoyant wall layer flow -- this is not seen, however, in the stratified case.}\label{snoplus31_effectof_stratification_A}
\end{figure}

To that end Fig.~(\ref{snoplus31_effectof_stratification_A}) examines the effect of stratification on the contaminant field in the wall boundary layer at time $t=8$ h, comparing two simulations that are both forced by $q=0.1 \, \mathrm{W \, m^{-2}}$ on the upper hemisphere wall. In one case, the initial state is isothermal ($T=290$ K), while in the other a uniform initial temperature gradient (of wall and fluid) was imposed, $T=290+0.1(z-5)$. The interesting point is that the stable stratification appears to have `immobilized' the wall boundary layer, since (contrary to the unstratified case) clean fluid has not moved upward in the wall layer to replace contaminated fluid. On this evidence alone, one might expect of the stratified case a lower rate of sink of the contamination layer.

\begin{figure}[h!]
\begin{center}
\includegraphics[width=14 cm]{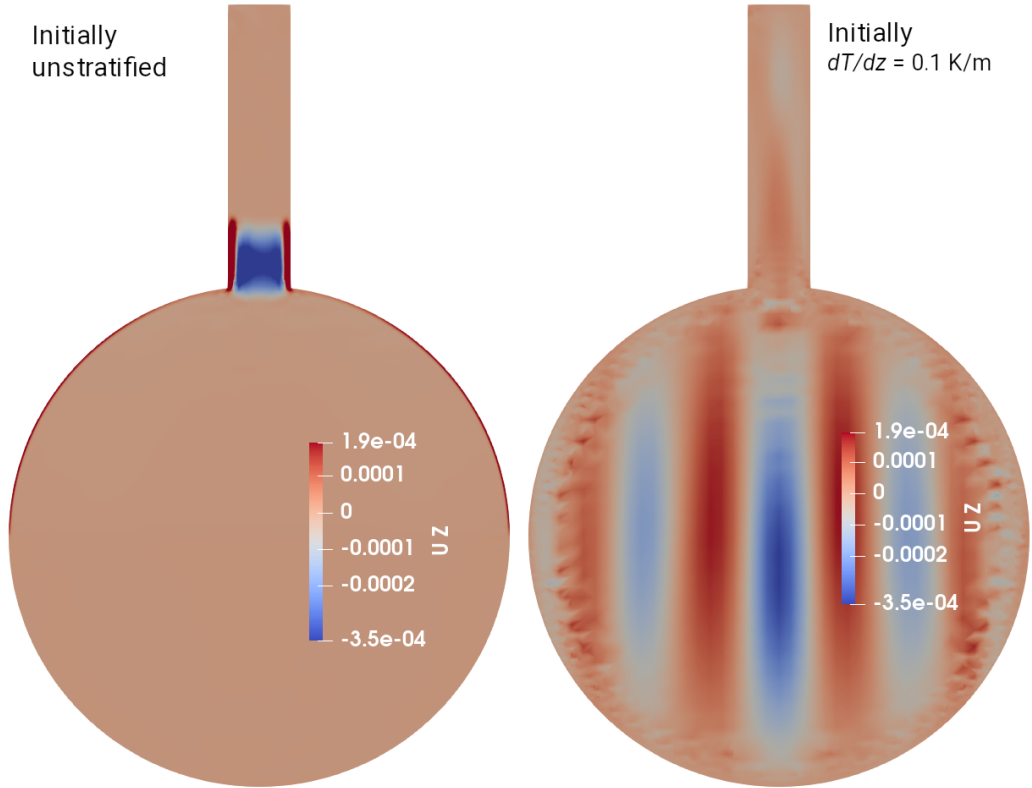}
\end{center}
\caption{Comparison of fields of vertical velocity UZ on $y=0$ at $t=8$ hr for two runs, both forced by heat flux $q=+0.1 \; \mathrm{W \, m^{-2}}$ on the upper hemisphere of the AV. Left: initial state isothermal. Right: initial state uniformly stratified, $T=290 + 0.1 (z-5)$. Stratification has diminished volume flux between the spherical AV and its neck, but resulted in an organised pattern of ascent/descent in the bulk of the AV. The colour scale on the right embraces the range in UZ (on $y=0$) for that case, whereas the range on the left has been fixed so as to match (the full range on the left was $-5.2 \times 10^{-4} \le \mathrm{UZ} \le 8.9 \times 10^{-4} \; \mathrm{m \, s^{-1}}$).}\label{snoplus31_effectof_stratification_B}
\end{figure}

\begin{figure}[h!]
\begin{center}
\includegraphics[width=13 cm]{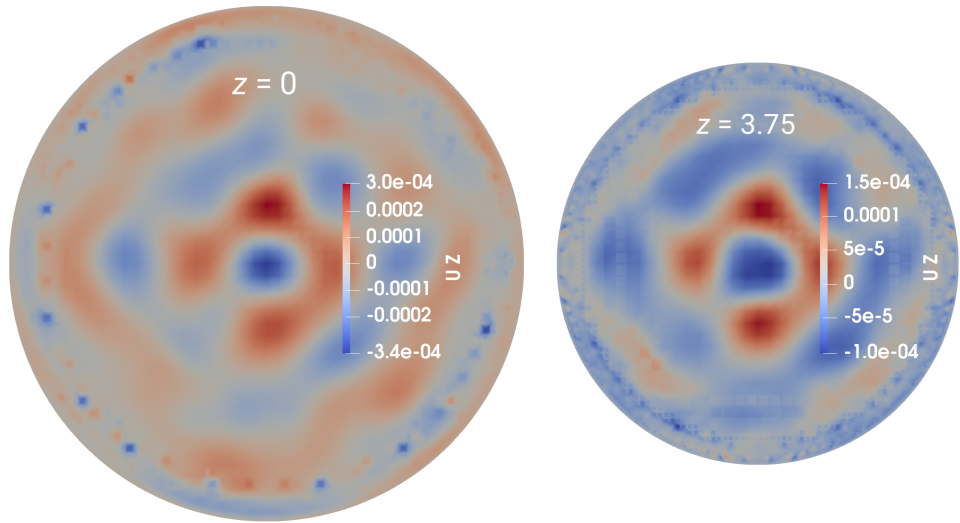}
\end{center}
\caption{Vertical velocity UZ on slices at $z=(0, 3.75)$ m at $t=8$ hr, for the simulation with stratified initial state and forced by heat flux $q=+0.1 \; \mathrm{W \, m^{-2}}$ on the upper hemisphere of the AV. Columns of ascent/descent cut these slices, showing vertical continuity (as also seen on Fig.~\ref{snoplus31_effectof_stratification_B}) and a far from perfect axial symmetry.}\label{snoplus31_effectof_stratification_C}
\end{figure}

\begin{figure}[h!]
\begin{center}
\includegraphics[width=17 cm]{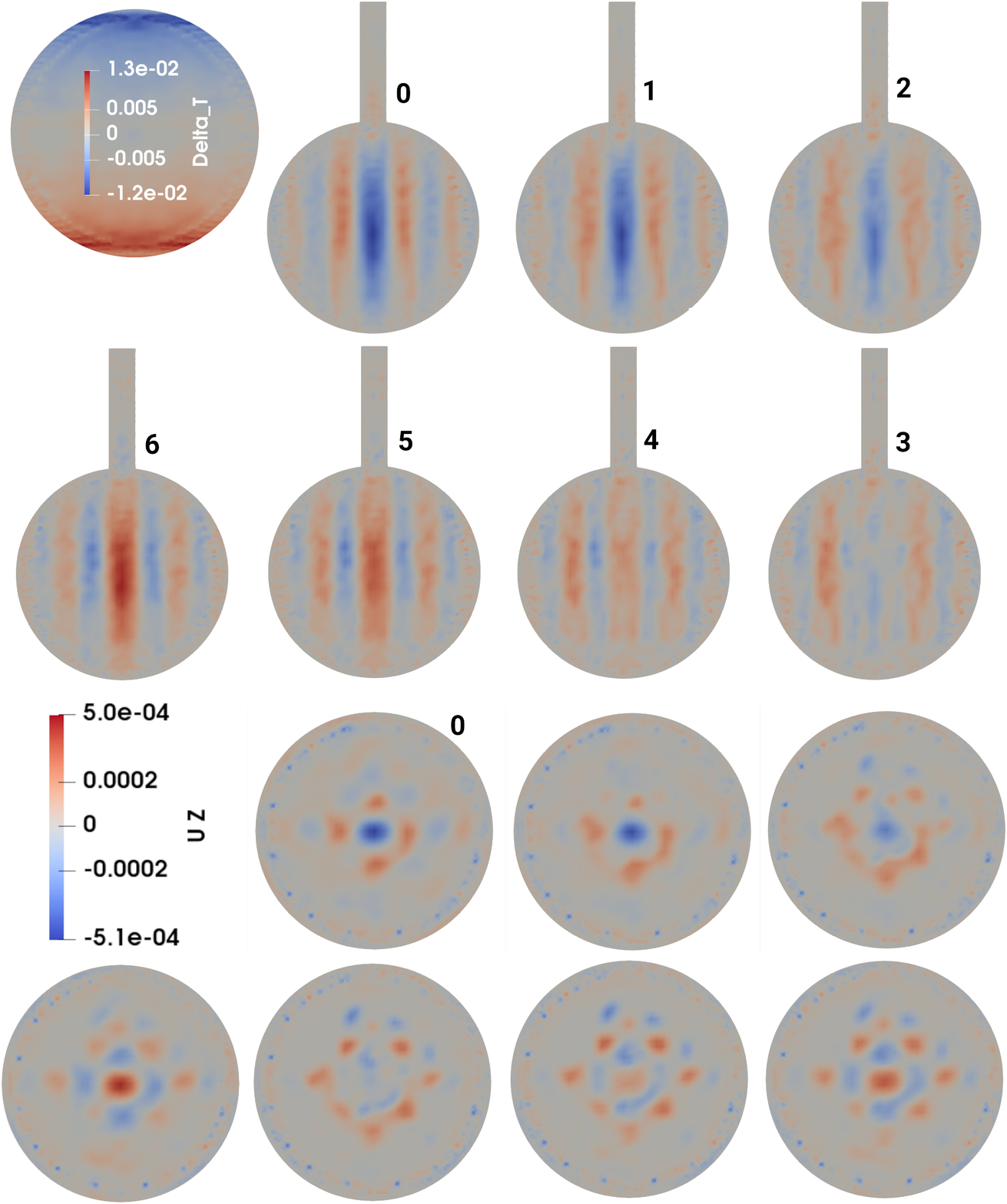}
\end{center}
\caption{Oscillatory pattern of vertical velocity UZ [$\mathrm{m \, s^{-1}}$], in stably-stratified detector forced by heat flux $q=+0.1 \; \mathrm{W \, m^{-2}}$ on the upper hemisphere of the AV. Vertical velocity sequence showing one half-cycle of oscillation, as viewed on a vertical slice ($y=0$) and an equatorial ($z=0$) slice. The interval between images (to be viewed clockwise, $0...6$) is $\Delta t=17$ s, and time `0' is $t=$7 hr + 17 s. The full period of the oscillation is approximately twelve steps (i.e. $204$ s), matching the Brunt-V\"{a}is\"{a}l\"{a} period (214 s). In the upper-left corner a spherical clip (radius 6 m) on $y=0$ shows temperature deviation [K] from the initial state ($T=290 + 0.1 (z-5)$) at $t=8$ hr. The colour scale for UZ is fixed across all plots.}\label{snoplus31_rn31Jan_oscillation}
\end{figure}

Fig.~(\ref{snoplus31_effectof_stratification_B}) compares the fields of vertical motion on $y=0$ for the same two simulations, and shows that stratification of the initial state has both reduced exchange between the neck and the spherical AV, and enhanced the organisation of flow in the latter, with the formation of a vertically coherent pattern of ascent and descent. This pattern, evident also on Fig.~(\ref{snoplus31_effectof_stratification_C}), shows roughly an axial symmetry. For this case the Brunt-V\"{a}is\"{a}l\"{a} (angular) frequency
\begin{equation}
\omega_{\mathrm{BV}} = \sqrt{g \, \beta_T \, \frac{dT_0}{dz}} \; ,
\end{equation}
based on the initial stratification (which essentially still prevails) gives a natural frequency $N_{\mathrm{BV}}=\omega_{\mathrm{BV}}/2\pi = 0.0047$ Hz, and a corresponding period $T_{\mathrm{BV}} = 214$ s. Fig.~(\ref{snoplus31_rn31Jan_oscillation}) gives a sequence of snapshots of vertical velocity at 17 s intervals, and establishes that the vertical motion pattern is {\em oscillatory} at the Brunt-V\"{a}is\"{a}l\"{a} frequency. Seen in animation, one can observe the axial columns migrate outward from the axis towards the AV wall\footnote{The horizontal velocity components, not shown, exhibit no obvious columnar organization, i.e. the wave motion does not feature organized axial rotation.}. A purely oscillatory motion can accomplish {\em no mixing}, which explains how it can be that the temperature field is little changed relative to the initial state (see panel on Fig.~\ref{snoplus31_rn31Jan_oscillation}) despite the vertical coherence of the velocity field.

\begin{figure}[h!]
\begin{center}
\includegraphics[width=9 cm]{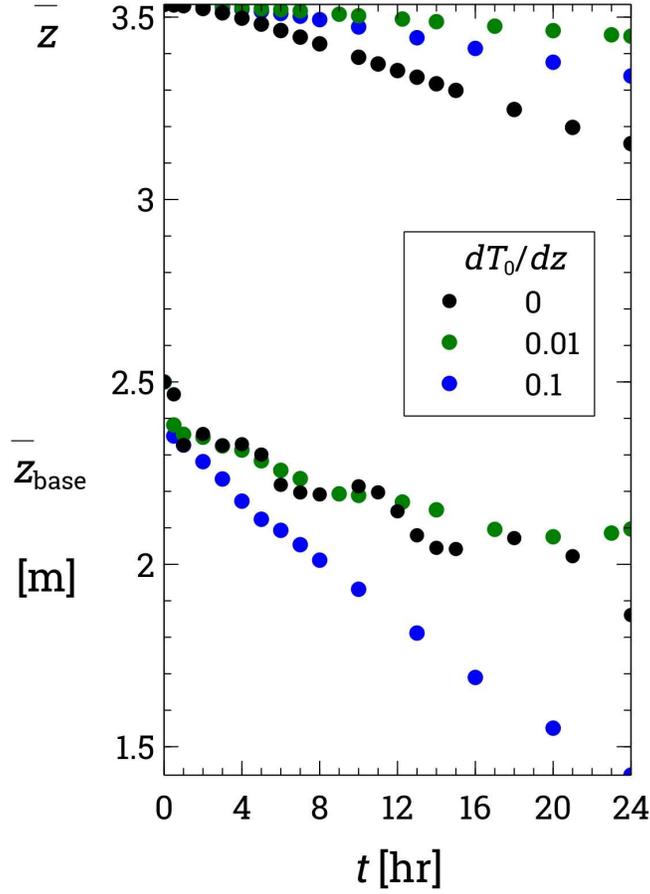}
\end{center}
\caption{Effect of initial thermal stratification $dT_0/dz$ on the rate of sink of a contamination layer, initially confined to $2.5 \le z \le 5$ m  where $C=1$. $\overline{z}$ is the contaminant-mass-mean height, and $\overline{z}_{\mathrm{base}}$ is the mean height over a subset of cells for which $z \le 2.5$ m and $0.05 f(t) \le C \le 0.1 f(t)$, where $f(t)=\exp(-t/\tau)$ is the radioactive decay factor, $\tau=(\ln 2)^{-1} \, T_{1/2}$ with $T_{1/2}=3.8$ days. Simulations forced by heat flux $q=+0.1 \; \mathrm{W \, m^{-2}}$ on the upper hemisphere of the AV.}\label{snoplus31_effectof_stratification_E}
\end{figure}

Returning to the effect of thermal stability on an idealized contamination layer (again, initially $C=1$ over $2.5 \le z \le 5$ m and $C=0$ elsewhere), Fig.~(\ref{snoplus31_effectof_stratification_E}) conveys an ambiguous picture. When sink rate is defined in terms of mass-mean contaminant height $\overline{z}$, weak and moderate stratification {\em reduce} the sink rate -- although the relationship between sink rate $d\overline{z}(t)/dt$ and initial stratification $dT_0/dz$ is not monotonic. Perhaps the complexity here relates to the fact that contaminant is also to some degree mixed {\em upward}: if bulk thermal stratification mitigates against ascent of the warm wall layer as suggested by Fig.~(\ref{snoplus31_effectof_stratification_A}), then perhaps some fluid-mechanical feedback, spontaneously arising in such a manner as to limit excessive heating of the near-wall fluid, accentuates upward mixing of the contaminant. But if we define a `base' height ($\overline{z}_{\mathrm{base}}$) for the contamination layer, as being the mean height of cells that (i) are below the initial base of the uniform contamination layer ($z < 2.5$ m) and (ii) carry `low' concentration\footnote{Arbitrarily $0.05 f(t) \le C \le 0.1 f(t)$, where $f(t)=\exp(-t/\tau)$ is the radioactive decay factor.}, Fig.~(\ref{snoplus31_effectof_stratification_E}) shows that firm stratification {\em enhances} descent of the base of the contamination layer. Admittedly this measure of base height is subjective and on the evidence of the figure somewhat erratic; but given that stable thermal stratification results (under the chosen forcing) in organized vertical motion in the bulk of the detector (Figs.~\ref{snoplus31_effectof_stratification_B}--\ref{snoplus31_rn31Jan_oscillation}), then an accelerated downward spread of the contamination layer within the spherical domain does not seem implausible.

\section{Contaminant transport down the neck}

The AV neck being the only path, aside from scintillator operations, by which radon is observed to enter the AV sphere, it is of interest to focus on flow and transport over that limited volume of the detector, i.e. a cylinder of radius 0.75 m and spanning $6 \le z \le 12.75$ m. Here it is relevant to clarify the thermal conditions prevailing {\em outside} the neck, as governed by its immersion in water over approximately the lower 6 m of its length but in cover gas over an uppermost section of about 1 m in length. The cover gas temperature is approximately that of the laboratory air, about $17 - 20^{\circ}$C, whereas the water is controlled at a considerably lower temperature of about $12^{\circ}$C by ``cavity recirculation''. The later process extracts water from the top of the column and returns it, after chilling, through two streams that respectively re-enter {\em within} and {\em outside} the PSUP structure, cooling water bodies referred to as ``PSUP water'' and ``cavity water''. It had been expected that this warm uppermost layer lying atop much cooler fluid would suppress convective motion in the neck, thereby limiting the rate of contaminant ingress and ensuring there would be at least a factor of 50 reduction in the $^{222}\mathrm{Rn}$- level in the scintillator at the bottom of the neck relative to the top.

Simulations of this section cover two simple and plausible forms of destabilizing thermal non-uniformity that induce convective mixing down the neck; more complex forms of thermal forcing are easily imagined, but those covered below suffice. The liquid/gas interface at $z=12.75$ m is treated as a `free slip/no leak' surface, and the sidewall and bottom boundary as `no slip/no leak' (i.e. $\mathbf{U}=0$).

\subsection{Steady wall heat flux}

Simulations of this sub-section were forced by a prescribed heat flux $q= \pm 0.1 \, \mathrm{W \, m^{-2}}$ on the side wall of the neck, disturbing an initial state that was either unstratified ($dT_0/dz=0$) or uniformly stratified ($dT_0/dz=0.025 \, \mathrm{K \, m^{-1}}$ or $0.25 \, \mathrm{K \, m^{-1}}$). The properties of this system imply a velocity scale
\begin{equation}
w_* = \left( \nu \, g \, \beta_T \, |q| \right)^{1/4} \; ,
\end{equation}
which for these simulations evaluates to $2.6 \times 10^{-4} \, \mathrm{m \, s^{-1}}$. Initially the contaminant concentration was $C=1$ for $z \ge 12.5$ m and $C=0$ elsewhere, although owing to mesh irregularity there was a degree of smearing across that nominal $C=1/0$ interface.

\begin{figure}[h!]
\begin{center}
\includegraphics[width=17 cm]{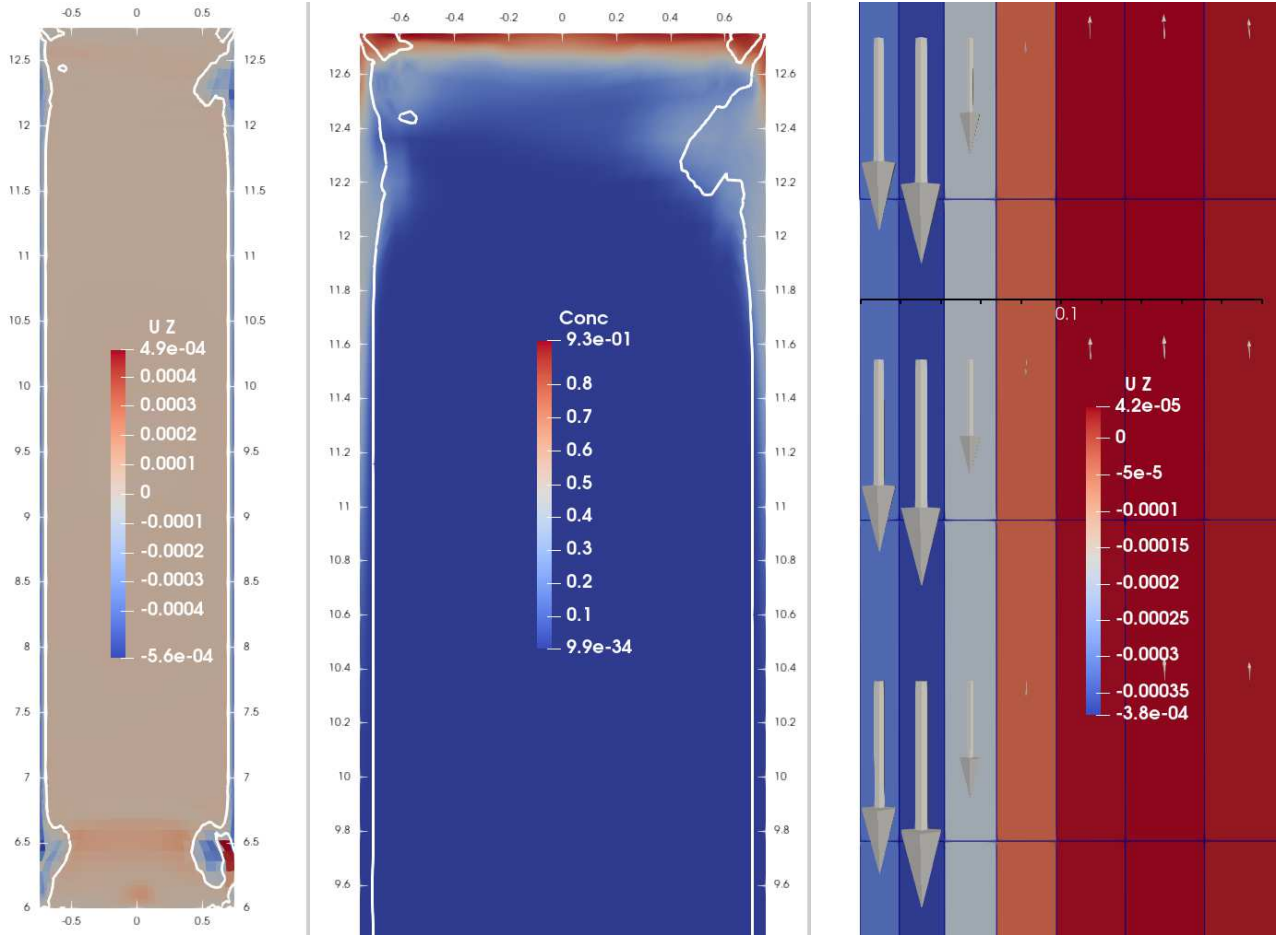}
\end{center}
\caption{Contaminant motion down the neck of the SNO+ AV. Slices on $y=0$, at $t=2$ h. Initial state isothermal, driven by $q=-0.1 \; \mathrm{W \, m^{-2}}$ on the neck sidewall. Each panel shows a segment (or all) of the $y=0$ plane, and in each case the colour scale range is adapted to the (whole or partial) view shown. The cool, sinking wall boundary layer (seen on left and right panels) is carrying a contaminant (`Conc', initially present with concentration $C=1$ at $z \ge 12.5$ m) downwards. The panel on the right gives a close-up view of the wall boundary layer, with the ruler (length 0.1 m, height $z=9.375$ m) indicating that at that point the boundary layer depth was about 0.05 m (and spanning 4-5 layers of cells). The single contour shown on the other two panels denotes $\mathrm{UZ}=0$, delineating regions of sinking and ascending fluid.}\label{snoplus34_run4Jan_t120min} 
\end{figure}

Fig.~(\ref{snoplus34_run4Jan_t120min}) shows the flow at $t=2$ h from initialization, when an isothermal initial state is forced by a cooling wall heat flux $q=-0.1 \, \mathrm{W \, m^{-2}}$. Sink in the wall boundary layer is necessarily compensated by weak ascent away from the walls, with the result that the contamination layer is drawn down along the wall, but elsewhere is compressed upwards. At $t=2$ h the standard deviation of the vertical velocity over the whole domain was $\sigma_{UZ} = 1.1 \times 10^{-4}  \, \mathrm {m \, s^{-1}}$, so that $\sigma_{UZ}/w_* = 0.4$. The largest magnitudes for vertical velocity were of order $5\times 10^{-4} \, \mathrm {m \, s^{-1}}$ (0.5 mm/s), two orders of magnitude larger than was the case in an {\em unforced} simulation at $t=2$ h.

\begin{figure}[h!]
\begin{center}
\includegraphics[width=17 cm]{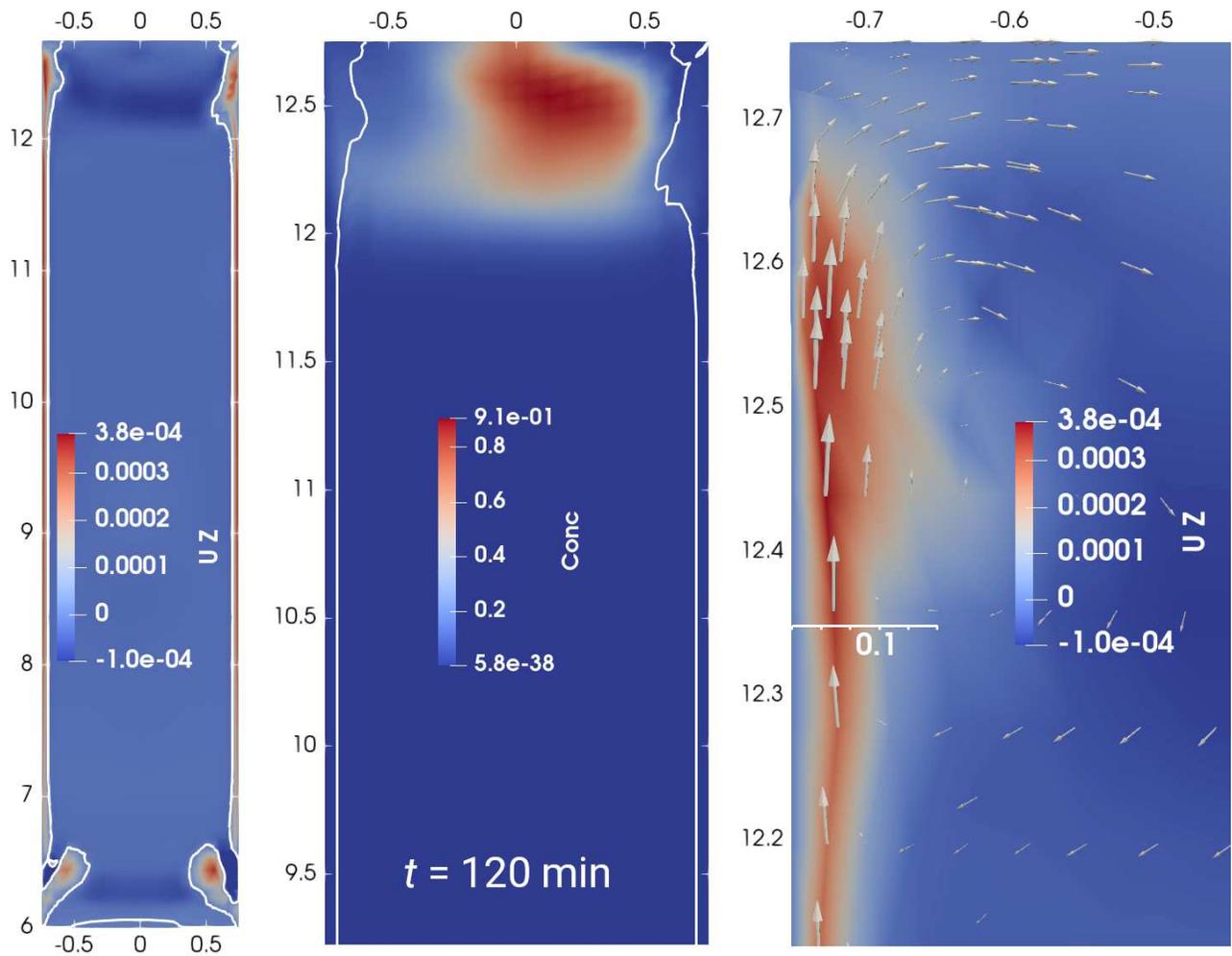}
\end{center}
\caption{Contaminant motion down the neck of the SNO+ AV. Slices on $y=0$, at $t=2$ h. Initial state isothermal, driven by $q = +0.1 \; \mathrm{W \, m^{-2}}$ on the neck sidewall. The single contour shown on the left and middle panels is $\mathrm{UZ}=0$. On the rightmost panel the ruler at $z=12.35$ m has length 0.1 m, and the vectors are proportional to $\mathbf{U}$ in magnitude and direction.}\label{snoplus34_run7Jan_t120min}
\end{figure}

Switching the sign of the wall heat flux, Fig.~(\ref{snoplus34_run7Jan_t120min}) shows a very different flow pattern at $t=2$ h, with the ascending volume flux of the wall boundary layer necessitating sink away from the wall. The contamination layer is as a result pushed away from the wall, and moves downward. The vertical velocity full scale range is very comparable with the previous case, and the statistic $\sigma_{UZ}/w_*$ is identical.

\begin{figure}[h!]
\begin{center}
\includegraphics[width=17 cm]{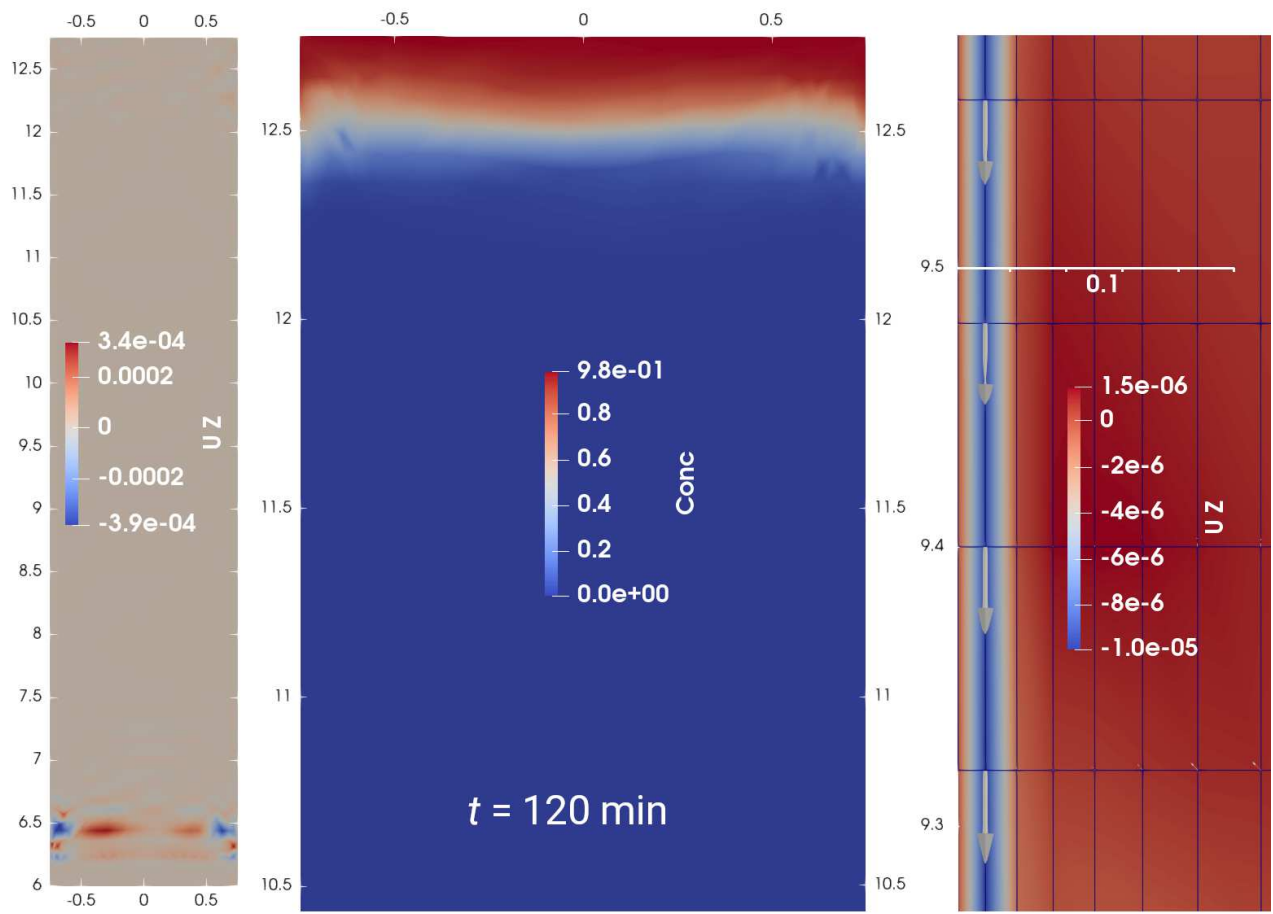}
\end{center}
\caption{Contaminant motion down the neck of the SNO+ AV. Slices on $y=0$, at $t=2$ h. Stably-stratified initial state $T=290+0.25 (z-6)$, driven by $q = - 0.1 \; \mathrm{W \, m^{-2}}$ on the neck sidewall. On the rightmost panel the ruler has length 0.1 m, and vectors are proportional to $\mathbf{U}$ in magnitude and direction. Though not evident from the figure, the cool wall boundary layer shows little variation in speed or depth over almost the entire length of the neck.}\label{snoplus34_run14Jan_t120min}
\end{figure}

When the initial state is thermally stratified, even weakly so, the simulations give evidence of wave motion, with the mesh size introducing a non-physical length scale on that pattern (see Fig.~\ref{snoplus34_run14Jan_t120min}). Whereas for the neutral cases a single closed $\mathrm{UZ}=0$ contour more or less partitioned the flow into ascent/descent regions, when initial stratification is imposed that contour degenerates into a maze of bubbles, and so is not shown. The stratification ($0.25 \; \mathrm{K \, m^{-1}}$) does not, however, impede the descent of the cool wall layer.

\begin{figure}[h!]
\begin{center}
\includegraphics[width=14 cm]{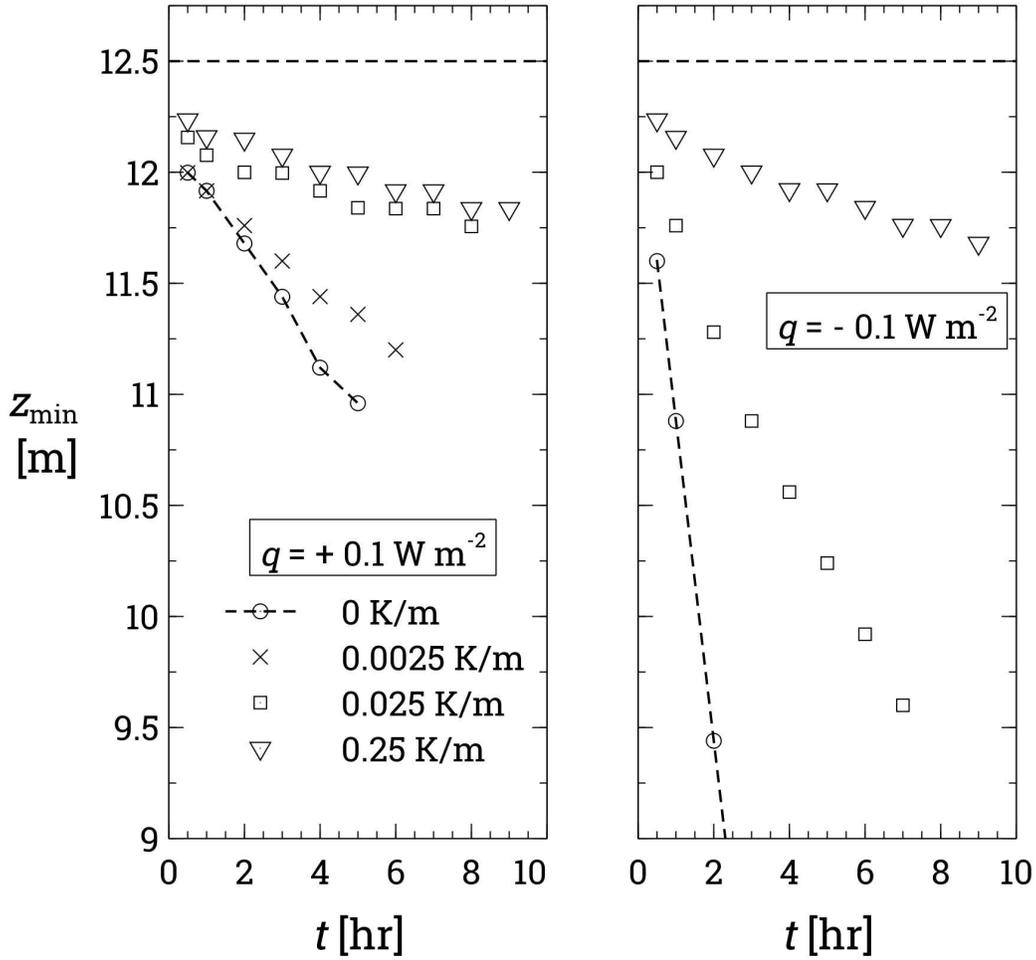}
\end{center}
\caption{Descent of a contamination layer (initially, $C=1$ at $z \ge 12.5$ m) down the neck of the SNO+ neutrino detector. Time evolution of the height ($z_{\mathrm{min}}$) of the lowest scalar-bearing cell ($C \ge 0.0005$) for various combinations of wall heat flux ($q= \pm 0.1 \; \mathrm{W \, m^{-2}}$) and initial stratification ($dT_0/dz, \; \mathrm{K \, m^{-1}}$). Note: $z_{\mathrm{min}}(0)$, not shown, is not {\em exactly} equal to 12.5 m (i.e. the nominal base of the initial contamination layer) due to the irregular shape and finite size of the mesh cells.}\label{snoplus34_zmin}
\end{figure}

Fig.~(\ref{snoplus34_zmin}) plots, for each of the simulations of this section, the height $z_{\mathrm{min}}(t)$ of the {\em lowest cell} bearing non-zero contaminant concentration, `non-zero' being arbitrarily taken to mean $C \ge 5 \times 10^{-4}$. The general effect of stratification is to retard the descent of fluid elements originating in the contamination layer. For the case $(q=-0.1, dT_0/dz=0)$ contamination reaches the base of the neck within about 4 hours, although the bulk of the contaminant remains trapped high in the neck in the weakly ascending flow away from the wall. For the situation $(q=+0.1, dT_0/dz=0)$ a longer time will be required before contamination first reaches the base of the neck, a point that must be qualified by noting that when the contaminant {\em does} first arrive, it will do so in greater volume, i.e. not only within the wall layer but in the bulk of the neck.

\subsection{Steady wall temperature anomaly}

In Section 6.1 motion in the neck was forced by a steady lateral flux of heat into the neck, which implies a continuous source of buoyancy --- that is, whatever the thermal response of the fluid adjacent to the wall, it continues to be undergo warming. In reality, and excepting unusual circumstances such as a failure of SNO+ water chillers, long intervals of steady heat addition or extraction from the SNO+ detector are not realistic. In this section simulations examine two instances wherein the motion is forced, instead, by a wall temperature anomaly relative to the initial fluid temperature, i.e. a {\em transient} wall temperature excess or deficit and thus a {\em transient} flux of buoyancy. The resulting adjustment of the fluid temperature to the (prescribed, constant) wall temperature implies that eventually a static steady (equilibrium) state can be expected.

In this section the origin for the $z$-axis is the base of the neck. The initial contaminant profile, irrelevant to the eventual steady state, was prescribed as linear ($C=z/7$). The upper and lower boundary conditions for tracer (i.e. radon) concentration were
\begin{equation*}
C = \begin{cases}
    1 \; , z = 7 \; \mathrm{m} \; , \\
    0 \; , z = 0 \; \mathrm{m} \;
    \end{cases}
\end{equation*}
and the sidewall boundary condition was specified as `zeroGradient' (i.e. no flux). At steady state this scenario must result in a downward flux of $C$ that, integrated over horizontal planes, must be independent of height. Furthermore $C_{\mathrm{avg}}(z)$, representing radon concentration averaged over the horizontal plane, serves as an indicative measure of the degree of `isolation' of any point in the neck relative to the contaminant load at the top of the neck.

\subsubsection{Transient negative buoyancy}

A simulation was performed with the neck idealized into three sections from base to top, with lengths $(L_1, L_2, L_3) = (3,3,1)$ m, and with the wall temperatures held fixed at respectively $(T_{1w}, T_{2w}, T_{3w}) = (12, 11.9, 20) \; ^{\circ}$C. This prescription was motivated by the fact of the uppermost section being surrounded by cover gas and the lower two sections by the much cooler cavity and PSUP water. Furthermore because the water masses (within and outside the PSUP sphere) are of non-uniform and time-varying temperature, destabilizing the system by holding the neck wall temperature of the middle section slightly colder than that of the lowest section seems within the range of possible circumstances. Initial {\em fluid} temperature was prescribed as
\begin{equation*}
T = \begin{cases}
    20^{\circ} \; , 6 < z \le 7 \; \mathrm{m} \; , \\
    12^{\circ} \; , z \le 6 \; \mathrm{m} \; .
    \end{cases}
\end{equation*}

\begin{figure}[h!]
\begin{center}
\includegraphics[width=15 cm]{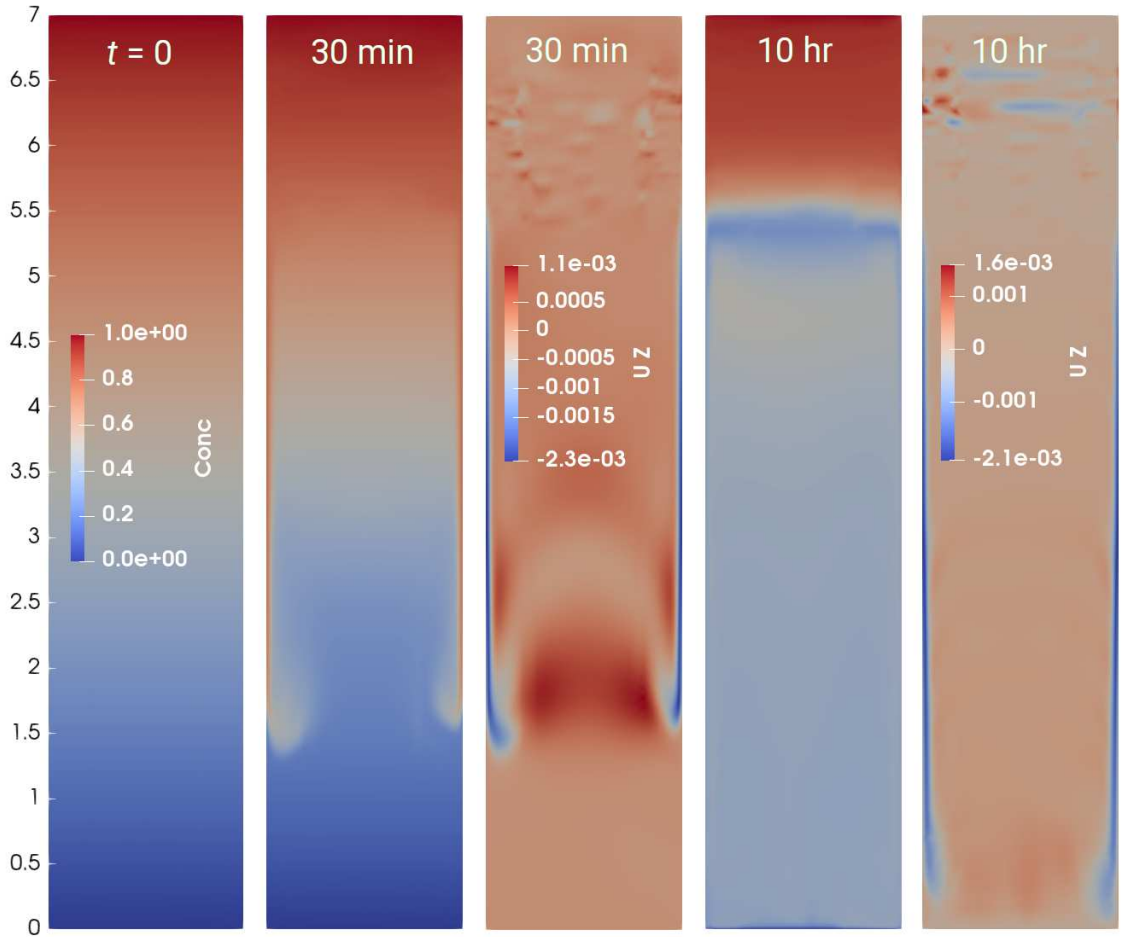}
\end{center}
\caption{Evolution of the contaminant field (`Conc' or $C$) and vertical velocity (UZ) in an idealization of the neck of the SNO+ neutrino detector, shown on $y=0$ slices. The motion is due to  inhomogeneity of wall temperature, which is fixed at $T_w=(12, 11.9,20)^{\circ}$ C in respectively the lowest 3 m, the central 3 m, and the uppermost 1 m of the neck. Initially (left panel) $C=z/7$, but that distribution (shown with same colour scale at $t=(30, 600)$ min) is quickly disturbed by the motion (though held by the boundary conditions to be 0/1 at base/top of the domain). By $t=30$ min the negatively-buoyant wall boundary layer is well developed, drawing contaminant downward. At $t=600$ min the contaminant distribution shows a sharp drop below the base of the uppermost (and sharply warmer) section.}\label{snoplus40_compos}
\end{figure}

\begin{figure}[h!]
\begin{center}
\includegraphics[width=12 cm]{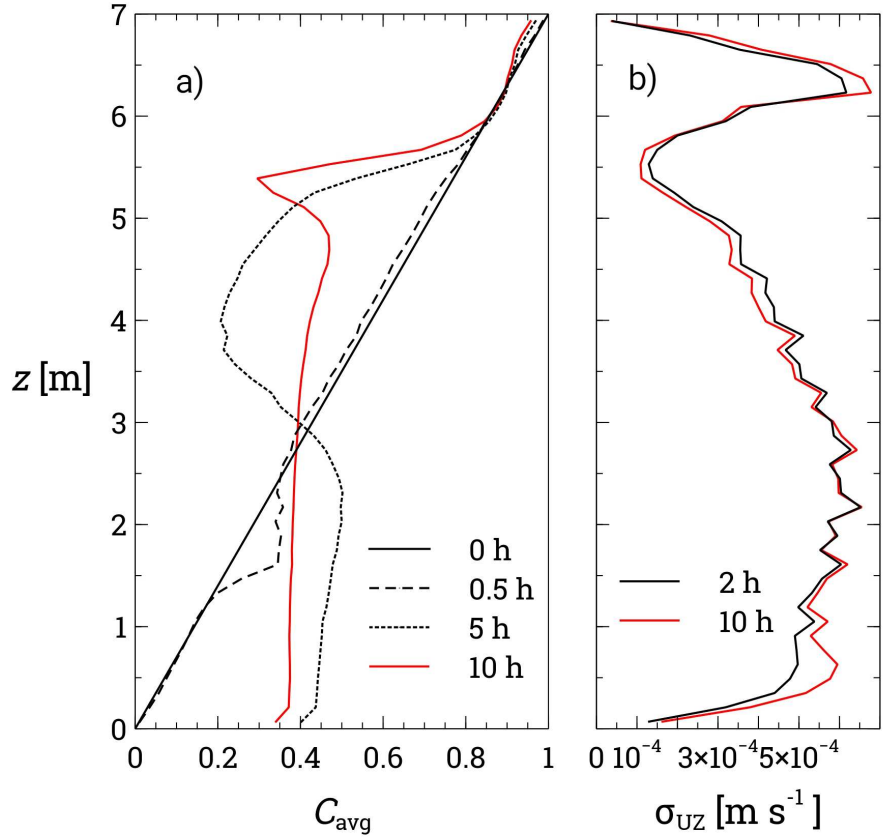}
\end{center}
\caption{Profiles of (a) horizontally-averaged concentration $C_{\mathrm{avg}}(z)$ and (b) the standard deviation $\sigma_{\mathrm{UZ}}(z)$ of vertical velocity. Properties defined over horizontal layers (of depth 0.14 m), in the simulation corresponding to Fig.~(\ref{snoplus40_compos}). Motion is the consequence of wall temperature in the layer $3 < z < 6$ m being slightly cooler than wall temperature at $z \le 3$ m.}\label{snoplus40_avg}
\end{figure}

Fig.~(\ref{snoplus40_compos}) summarizes the evolution of the concentration and vertical velocity fields to $t=10$ hr. A negatively buoyant wall boundary layer develops rapidly, and draws contaminant downwards along the neck, while the compensating vertical motion outside the boundary layer lifts tracer upwards. By 10 hr the contaminant field has evolved (roughly speaking) into two well mixed layers, with the boundary between the upper and lower layers of respectively high and low concentration standing about 0.5 m below the base of the warm wall segment. The profile of horizontally-averaged concentration ($C_{\mathrm{avg}}$, see Fig.~\ref{snoplus40_avg}a) captures this `step' between high concentration above and somewhat lower concentration below the base of the warm layer, and the cause of the step is explicit in the profile of the standard deviation of vertical velocity ($\sigma_{\mathrm{UZ}}$, see Fig.~\ref{snoplus40_avg}b): vertical motion is nulled not only at the top and bottom boundaries (courtesy of the imposed boundary condition), but also greatly suppressed near the gravitationally stable temperature step. If $C_{\mathrm{avg}}$ is taken as an indication of the extent to which contaminant concentration is buffered relative to its value at the top of the neck, then this simulation indicates that the thermal stability owing to existence of a much warmer fluid layer at the top of the neck cannot suppress the vertical exchange that owes to a wall temperature inhomogeneity even so small as of $\mathcal{O}[0.1 \, \mathrm{K}]$, for $C_{\mathrm{avg}}$ remains $\mathcal{O}[1]$ in the lower layers, excepting at the base (where it is artifically held to $C=0$ by the boundary condition).

\subsubsection{Transient positive buoyancy}

The final simulations were motivated by the intuition that strong (stable) stratification of the detector ought to suppress or at least limit the mixing induced by transient and weak thermal disturbances. The initial fluid temperature profile was specified as
\begin{equation*}
T = 285 + \gamma \, z
\end{equation*}
with $\gamma = 1 \; \mathrm{K \, m^{-1}}$, and the (steady) wall temperature as
\begin{equation*}
T_w = 285 + \gamma \, z + \Delta T \; \exp \left[- \frac{(z-1)^2}{2 \sigma^2} \right]
\end{equation*}
with (for the case to be shown) $\Delta T=4$ K and $\sigma=0.4$ m.

\begin{figure}[h!]
\begin{center}
\includegraphics[width=15 cm]{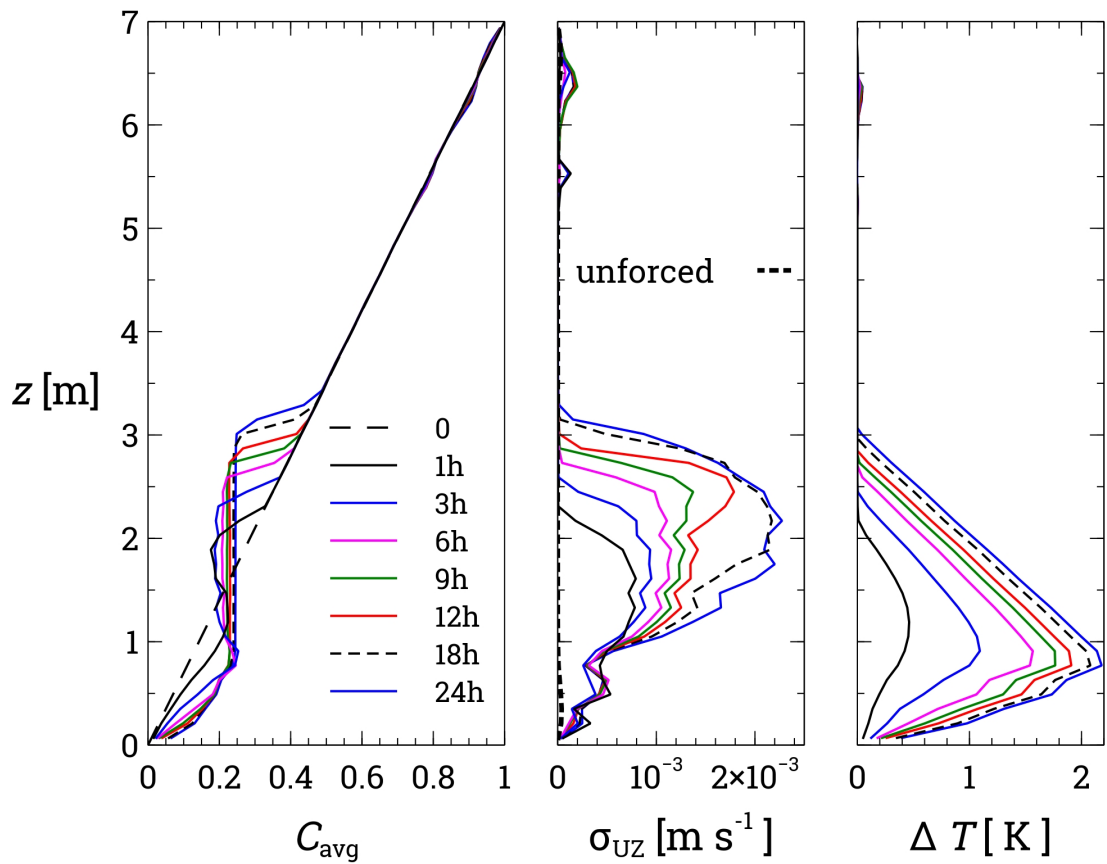}
\end{center}
\caption{Profiles of (a) horizontally-averaged concentration $C_{\mathrm{avg}}(z)$, (b) standard deviation $\sigma_{\mathrm{UZ}}(z)$ of vertical velocity and (c) temperature rise relative to the initial state. Properties for each time are defined over horizontal layers (of depth 0.14 m). Motion is the consequence of a `Gaussian' warm ring of wall temperature centred at $z=1$ m, where the excess temperature is 4 K. The `unforced' profile of $\sigma_{\mathrm{UZ}}$, quantifying the level of false (or `numeric') mixing, is for $t=9$ hr from a simulation on the same mesh and differing only in that there is no forcing warm wall ring (the level of unforced motion is to all intents and purposes time-independent).}\label{snoplus40_rn10Oct_avg}
\end{figure}

Fig.~(\ref{snoplus40_rn10Oct_avg}) summarizes the simulation, which reaches a steady state by about $t=18$ h. The consequence of the ring of warm wall temperature is a ``mixed layer'' within which the scalar contaminant ($C_{\mathrm{avg}}$) is uniform. As intuition would suggest, the forcing results in mixing that is restricted to a layer over which heated parcels rise due to their buoyancy relative to the cooler fluid about them\footnote{Having attained an upward momentum, parcels are liable to ``overshoot'' and rise above their layer of neutral buoyancy, decelerate, then sink again.}. Also evident on Fig.~(\ref{snoplus40_rn10Oct_avg}) is a degree of `false' or `numeric' motion and mixing, most clearly seen around $z=6.5$ m. Characterized by $\sigma_{\mathrm{UZ}}$ the level of unforced motion quickly comes to a steady state, and is certainly an artiface of the numerical solution: in unforced simulations $\sigma_{\mathrm{UZ}}$ diminishes with increasing mesh refinement. It owes to discretization errors which are larger in regions where the mesh cells (mostly hexahedra) feature non-orthogonal boundaries, such as as near intersecting domain boundaries that are endowed with `added' cell layers.

Equivalent calculations with a weaker wall temperature excess produce a correspondingly narrower mixed layer, and one can conclude (without surprise) that in combination with uniform neck stratification $\gamma \, [\mathrm{K \, m^{-1}}]$ a modest wall temperature anomaly of magnitude $T'$ will induce (transient) mixing over a depth of order $\gamma^{-1} \, T'$. Presumably this simple argument has its limits -- for example if the perturbation were to induce a motion field sufficiently strong as to erase the initial stratification.

\section{Discussion and Conclusion}

Apart from numerous geophysical and planetary studies, the literature on convective flow of a materially homogeneous fluid {\em within} a sphere is surprisingly sparse. The present study of motion in the SNO+ liquid scintillator neutrino detector has been motivated by what had been a puzzling phenomenon, and was initiated with the goal of providing a plausible explanation for that motion. The numerical method (OF's standard solver `buoyantPimpleFoam') being well tried, we have relied on the demonstrated consistency of simulations with measurements in a geometrically congruent small scale experiment (Section \ref{sec:ChowAkins}) as testimony to the realism -- qualitative or better -- of the full scale (SNO+) calculations.

The simulations suggest that the phenomenon prompting the study, i.e. the slow descent within the detector of a radon contamination layer, was driven by volume exchange from beneath to above that layer, the needed upward volume flux ($Q_{\mathrm{vol}}^{+}$) taking place due to and within a thin convective boundary layer on the AV wall. Consistency of the simulation with the observed multi-day phenomenon hinges to some extent on the absence of any firm constraint in specifying the {\em forcing} for the simulated flow, but our choice, viz. a weak heat flux density $q=0.1 \, \mathrm{W \, m^{-2}}$ into the fluid from the wall of the upper hemisphere, is not dissimilar from what is implied by (sporadically observed) thermal expansion of the scintillator fluid. Interestingly too, short episodes of contamination {\em rising and spreading laterally} from the base of the detector, seen after intervals of so-called `reverse' circulation of the scintillator fluid -- `reverse' meaning the LAB is reintroduced at the base of the detector, rather than the base of the neck -- are consistent with OF simulations that impose a cooling heat flux around the {\em lower} hemisphere of the detector.

The azimuthal symmetry of the SNO+ detector may suggest to readers that the full 3-dimensional (3D) treatment adopted here was uneconomic, and certainly theoretical studies of convection within a sphere (e.g. \citep{Hutchins_Marschall1989,Whitley_Vachon1972,AnguianoOrozco_Avila}) have presupposed azimuthal symmetry. Over the course of the work some early simulations adopted as the domain just a single quadrant of the detector, but the additional planar boundaries made mesh generation a more complex task; and as to reducing the problem to 2D, although this would be fine in the context of steady state treatment, in 2D one loses temporal fidelity. In that regard, it appears that there are two or more timescales for the development of this flow. The shortest timescale $\tau_{\mathrm{bl}}$ ($\sim 1 \; \mathrm{hr}$, or smaller) relates to the establishment of motion in the wall boundary layer, and so soon as total time $t \gg \tau_{bl}$ (i.e. a few hours after initiation) the sink rate of the blob is established, albeit decreasing with increasing $t$ as the blob sinks, with associated increase of its planar surface area. On a longer time scale ($\tau_{\mathrm{bulk}}$) gross thermal stratification (in response to the wall heat flux) and quasi-horizontal swirl (presumably originating in the Coriois term) are established in the bulk of the AV.

Overall this work indicates that even what might seem to be `small' thermal inhomogeneities or trends can suffice to bring about significant motion (up to $\mathcal{O}[\mathrm{mm/sec}]$) in this type of fluid `body', and the fact of these simulations being for a spherical geometry probably does not limit this suggestion to that particular shape. The motion engendered by {\em some} forms of thermal disturbance can be mitigated by ensuring that a stabilizing vertical thermal gradient is sustained, but in this regard the {\em duration} and {\em strength} of the forcing disturbance are important -- a sustained buoyancy flux is liable, eventually, to induce complete mixing. In short, if scintillator motion is to be avoided then (horizontal) thermal homogeneity must be assured to a very fine level.

\subsection*{Acknowledgements}

The author thanks the SNO+ collaboration for access to the data that spurred the questions addressed here, and in particular Drs. A. Wright, V. Lozza, A. Hallin and C. Krauss, who provided regular suggestions and feedback that guided the progression of this work. Dr. J.-S. Wang produced the SNO+ data plots shown in Figs.~(2,5). Funding for the author's participation in SNO+ derives from a Natural Sciences and Engineering Research Council of Canada (NSERC) Discovery Grant.

\setlength{\baselineskip}{12pt}

\bibliographystyle{elsarticle-num}
\bibliography{NIMreferences}

\pagebreak 

\section*{TABLES}

\begin{table}[htbp!]
\caption{The `thermoType' block in the OpenFoam dictionary (file) ``thermophysicalProperties'', as used for simulations of the Chow-Akins experiment and of the SNO+ detector.}\label{tab:thermoType_in_thermophysicalProperties}
\begin{Verbatim}[frame=single]
thermoType
{
    type            heRhoThermo;
    mixture         pureMixture;
    transport       polynomial;
    thermo          hPolynomial;
    equationOfState icoPolynomial;
    specie          specie;
    energy          sensibleEnthalpy;
} 
\end{Verbatim}
\end{table}
\normalsize \setlength{\baselineskip}{18pt}

\begin{table}[htbp!]
\caption{Specification (in standard OF file `casefolder/0/T') of the initial and boundary conditions on temperature, for simulation of the Chow-Akins experiment. An integer identifier (arbitrarily named `origin\_cellID') is associated with the coordinate origin (i.e. centre of the sphere). The arbitrarily named variable `T00\_plus' is set equal to the temperature at cell `origin\_cellID' incremented by 2.5 K, and imposed as the temperature everywhere on the domain boundary (whose patch name is `shell'). Thus the domain boundary is sustained at a temperaure 2.5 K warmer than the temperature at the origin.}\label{tab:init_and_bconds_ChowAkins}
\begin{Verbatim}[frame=single]
internalField   uniform 278.0;

boundaryField
{
    shell
    {
        type            codedFixedValue;
        value           uniform 278;
        name            T00_plus;
        code            #{            
          const volScalarField& T = db().lookupObject<volScalarField>("T");
          const fvMesh & mesh = T.mesh();                    
          scalar origin_cellID = mesh.findCell(point(0.0, 0.0, 0.0));            
          scalar value = T[origin_cellID]+2.5; // wall temperature excess
          operator==( value );           
        #};
    }
} 
\end{Verbatim}
\end{table}
\normalsize \setlength{\baselineskip}{18pt}

\begin{table}[htbp!]
\caption{Values used for the material properties of LAB. Note: Trials with a wide range of alternative values for $\mathrm{Sc}$ suggested the uncertainty in its value is (for present purposes) inconsequential.}\label{tab:properties}
\begin{tabular}{l l l}
\\ \hline\\
{\bf Property}                            & {\bf Value}                                    & {\bf Source}\\
\hline 
Density, $\rho_0$                         & $858 \, \mathrm{kg \, m^{-3}}$                 & Zhou et al. \cite{Zhou_etal_2015}) \\
Kinematic viscosity, $\nu$                & $1\times10^{-5}\,\mathrm{m^2\,s^{-1}}$         &  Material Data Safety Sheet (``5-10 cps'') \\
Thermal expansion coefft., $\beta_T$      & $8.8\times10^{-4}\,\mathrm{K^{-1}}$            & Zhou et al. \cite{Zhou_etal_2015}) \\
Thermal diffusivity., $\kappa$            & $7.2\times10^{-8}\,\mathrm{m^2 \, s^{-1}}$     & Wu et al. \cite{Wu_etal2019})\\
Specific heat capacity, $c_p$             & $2300 \, \mathrm{J \, kg^{-1} \, K^{-1}}$      & Wu et al. \cite{Wu_etal2019}) \\
Prandtl no., $\mathrm{Pr}$(=$\nu/\kappa$) & 140                                            & (rounded)\\
Schmidt no., $\mathrm{Sc}$                & 140                                            & (surmised) \\
\\ \hline
\end{tabular}
\end{table}
\normalsize \setlength{\baselineskip}{18pt}

\begin{table}[htbp!]
\caption{The `mixture' block in the OpenFoam dictionary (file) ``thermophysicalProperties'', for SNO+ simulations. Note: `mu' ($\mu$) is the {\em dynamic} viscosity ($\mu = \rho \nu$) and `kappa' is the thermal conductivity ($\kappa, \, \mathrm{W \, m^{-1} \, K^{-1}}$). The reference temperature $T_0$ for the polynomial giving the density, $\rho=\rho_0 + \beta_T \, (T-T_0)$, has been taken as 290 K.}\label{tab:mixture_in_thermophysicalProperties_SNO+}
\begin{Verbatim}[frame=single]
mixture
{
    specie
    {
        molWeight   240.0;
    }
    thermodynamics                                                             
    {                                                                          
        CpCoeffs<8>     (2300.0 0 0 0 0 0 0 0);
        Sf              0;                                                     
        Hf              0;                                                     
    }                                                                          
    equationOfState                                                       
    { 
        rhoCoeffs<8>    (1077 -0.755 0 0 0 0 0 0);
    }                                                                          
    transport                                                                  
    {                                                                          
        muCoeffs<8>     (0.00858 0 0 0 0 0 0 0);        
        kappaCoeffs<8>  (0.143   0 0 0 0 0 0 0);
        Pr              140;        
    }                                                                          
} 
\end{Verbatim}
\end{table}
\normalsize \setlength{\baselineskip}{18pt}

\clearpage \pagebreak

\section*{FIGURES}

\end{document}